\documentclass[twocolumn,twocolappendix]{aastex7} 
\usepackage{hyperref}
\usepackage{amsmath,amstext}
\usepackage[T1]{fontenc}
\usepackage{graphicx}
\usepackage[figure,figure*]{hypcap}
\usepackage{multirow}
\usepackage{comment}
\usepackage{nicefrac}

\usepackage{stackengine}

\usepackage[]{algorithm2e} 

\usepackage{lineno}

\def\SPARK{\texttt{SPARK}}
\def\TARDIS{\texttt{TARDIS}}
\def\approxposterior{\texttt{approxposterior}}

\usepackage{soul}

\defcitealias{vieira23}{V23}
\def\vieiratwothree{\citetalias{vieira23}}

\defcitealias{vieira24}{V24}
\def\vieiratwofour{\citetalias{vieira24}}

\def\sun{$_{\odot}$}

\def\f28{${f}_{2-8{\rm keV}}$}

\def\sun{$_{\odot}$}

\def\ergscm2{erg s$^{-1}$ cm$^{-2}$}
\def\yr-1{yr$^{-1}$}

\def\eg{{\it e.g.}}

\def\ie{{\it i.e.}}

\def\a{$\&$}
\def\x{$\times$}

\def\w{$\omega$}

\def\asec{\ifmmode^{\prime\prime}\else$^{\prime\prime}$\fi}

\hypersetup{linkcolor=magenta}

\shorttitle{Linking Spectral and Light Curve Modeling of the GW170817 Kilonova} 
\shortauthors{Vieira {\it et al.}}

\begin{document}

\title{Spectroscopic $r$-Process Abundance Retrieval for Kilonovae III: Linking Spectral and Light Curve Modeling of the GW170817 Kilonova}

\correspondingauthor{Nicholas~Vieira}
\email{nicholas.vieira@mail.mcgill.ca}

\author[0000-0001-7815-7604]{Nicholas~Vieira}
\affil{Trottier Space Institute at McGill, McGill University, 3550 rue University, Montreal, Qu{\'e}bec, H3A 2A7, Canada}
\affil{Department of Physics, McGill University, 3600 rue University, Montreal, Qu{\'e}bec, H3A 2T8, Canada}
\email{nicholas.vieira@mail.mcgill.ca}

\author[0000-0001-8665-5523]{John~J.~Ruan}
\affil{Department of Physics and Astronomy, Bishop's University, 2600 rue College, Sherbrooke, Qu{\'e}bec, J1M 1Z7, Canada}
\email{jruan@ubishops.ca}

\author[0000-0001-6803-2138]{Daryl Haggard}
\affil{Trottier Space Institute at McGill, McGill University, 3550 rue University, Montreal, Qu{\'e}bec, H3A 2A7, Canada}
\affil{Department of Physics, McGill University, 3600 rue University, Montreal, Qu{\'e}bec, H3A 2T8, Canada}
\email{daryl.haggard@mcgill.ca}

\author[0000-0001-7081-0082]{Maria~R.~Drout}
\affil{David A. Dunlap Department of Astronomy and Astrophysics, University of Toronto, 50 St. George St., Toronto, Ontario, M5S 3H4, Canada}
\email{maria.drout@utoronto.ca}

\author[0000-0003-4619-339X]{Rodrigo Fern{\'a}ndez}
\affil{Department of Physics, University of Alberta, Edmonton, Alberta, T6G 2E1, Canada}
\email{rafernan@ualberta.ca}

\begin{abstract}

The observed spectra and light curves of the kilonova produced by the GW170817 binary neutron star merger provide complementary insights, but modeling both the spectral- and time-domain has proven challenging. Here, we model the optical/infrared light curves of the GW170817 kilonova, using the properties and physical conditions of the ejecta as inferred from detailed modeling of its spectra. Using our software tool \texttt{SPARK}, we first infer the $r$-process abundance pattern of the kilonova ejecta from spectra obtained at 1.4, 2.4, 3.4, and 4.4 days post-merger. From these abundances, we compute time-dependent radioactive heating rates and the wavelength-, time-, and velocity-dependent opacities of the ejecta. We use these inferred heating rates and opacities to inform a kilonova light curve model, to reproduce the observed early-time light curves and to infer a total ejecta mass of $M_{\mathrm{ej}} = {0.11}~M_{\odot}$, towards the higher end of that inferred from previous studies. The combination of a large ejecta mass from our light curve modeling and the presence of both red and blue ejecta from our spectral modeling suggests the existence of a highly magnetized hypermassive neutron star remnant that survives for $\sim$$0.01 - 0.5$~s and launches a blue wind, followed by fast, red neutron-rich winds launched from a magnetized accretion disk. By modeling both spectra and light curves together, we demonstrate how combining information from both the spectral and time domains can more robustly determine the physical origins of the ejected material.

\end{abstract}


\section{Introduction}\label{sec:intro}

\begin{figure*}[!ht]
    \centering
    \includegraphics[width=1.\textwidth]{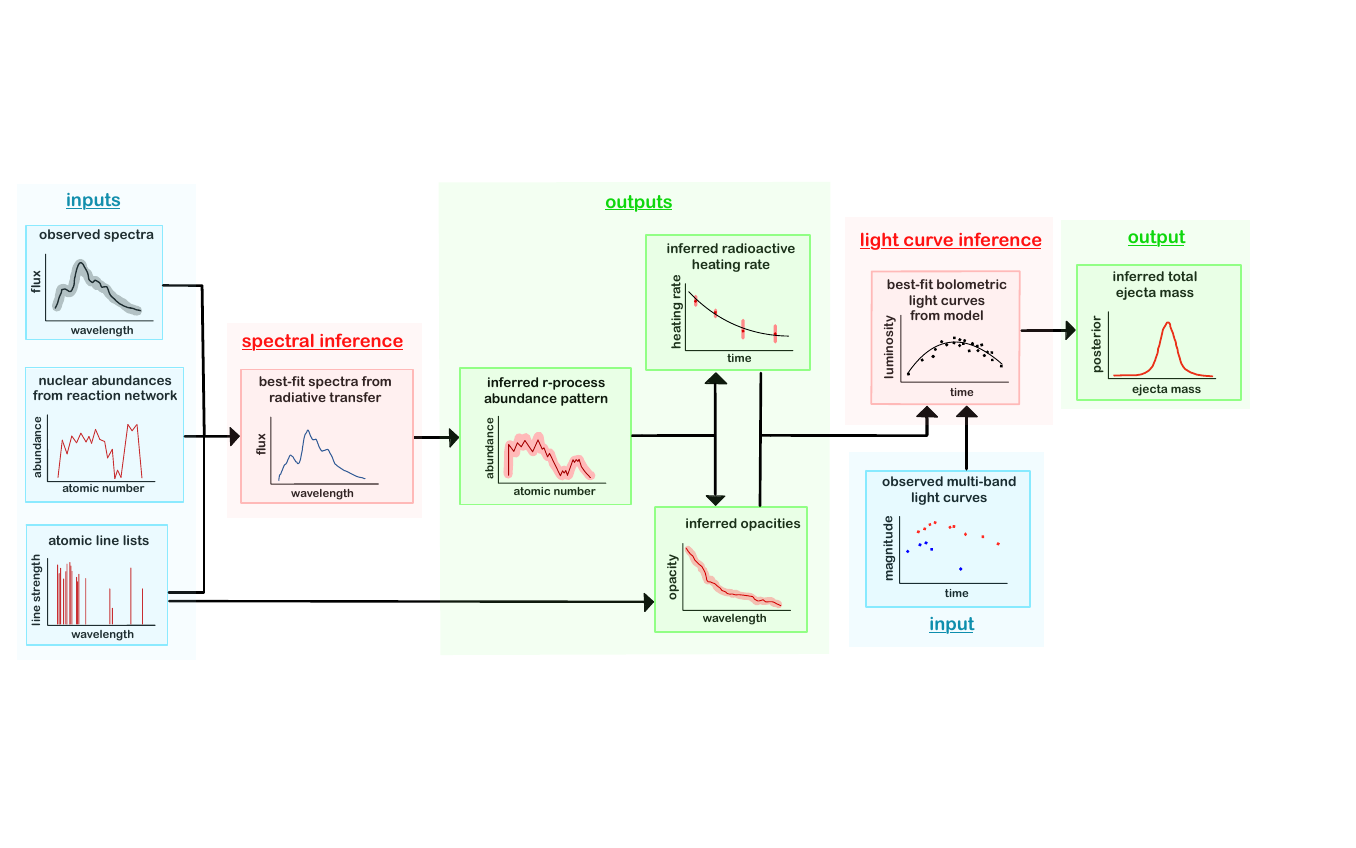}
    \caption{\textbf{Schematic diagram illustrating our spectral and light curve inference on observations of the GW170817 kilonova.} We input a grid of abundance patterns from nuclear reaction network calculations and lab-measured atomic line lists into radiative transfer simulations to fit the observed optical/IR spectra of the GW170817 kilonova using our tool~\texttt{SPARK}. This spectral modeling enables Bayesian inference of the abundance pattern of $r$-process elements synthesized in the ejecta. From these inferred abundance patterns, we compute the heating rates and opacities of the ejecta, which we then input into a light curve model. We fit the bolometric light curve estimated from the observed multi-band light curves of the kilonova to infer a total ejecta mass. This procedure combines both spectral- and time-domain information, ensuring that our spectral and light curve analyses make minimal assumptions on key parameters such as abundances, heating rates, and opacities.}
    \label{fig:schematic_lightcurves}
\end{figure*}

The landmark binary neutron star (BNS) merger GW170817 is the only event observed in both gravitational waves and electromagnetic radiation (\citealt{abbottLIGO17_MM}), and is among the most intensely studied events in astrophysics. The resultant kilonova transient was detected in the ultraviolet (UV)-optical-infrared (IR) and is well-characterized by a rich dataset, spanning $\sim$10 hours to weeks post-merger. BNS merger ejecta contain a high density of free neutrons that are conducive to rapid neutron capture (the $r$-process), which synthesizes about half of all elements heavier than iron (\citealt{cowan21}). Radioactive isotopes of these freshly-synthesized elements decay, and the energy from these decays thermalizes in the ejecta, thus powering the kilonova transient. 

Analysis of the GW170817 kilonova has relied primarily on radiative transfer modeling (\eg, \citealt{kasen17, chornock17, watson19, domoto22, gillanders22, gillanders24, sneppen24}) and analytic/semi-analytic models (\eg, \citealt{cowperthwaite17, drout17, kasliwal17, villar17, rosswog18, waxman18, hotokezaka20}). Analytic treatments have been especially common in studying the multi-band light curves. An important ingredient in light curve models is the radioactive heating rate, $\dot{q}(t)$. This $\dot{q}$ (energy per unit time per unit mass) describes how much energy is produced by radioactive decays, which is then thermalized according to some thermalization efficiency $\epsilon_{\mathrm{th}}(t)$ (\eg, \citealt{barnes16, wollaeger18}), dictating the amount of energy deposited into the ejecta. A common prescription for $\dot{q}$ is the semi-analytic fit to the heating rates from the nuclear reaction network calculations of \cite{korobkin12}. However, as pointed out in \cite{sarin24b} and elsewhere, this prescription is most appropriate for ejecta with low electron fractions (which trace the neutron richness of the ejecta) of $Y_e \approx 0.05$. For this low $Y_e$, the ejecta is highly neutron-rich and contains a suite of isotopes of varying half-lives that combine to produce a power-law heating rate. \cite{rosswog24} improve upon this $\dot{q}$ with a more broadly-applicable prescription that extends to higher $Y_e$, where the heating rate may be dominated by individual isotopes, thus deviating from a power-law. 

Another important ingredient in light curve modeling is the opacity of the ejecta. Typically, models assume a `gray' (wavelength-independent) opacity $\kappa$, with the specific value adopted dependent on the composition of the ejecta.  Lanthanides, which are produced at sufficiently low $Y_e$, have higher opacities due to their wealth of line transitions, especially in the UV and optical. For example, \cite{tanaka20} provide estimates of the opacity for different $Y_e$ at a density $\rho = 10^{-13}~\mathrm{g~cm^{-3}}$ and temperature $T = 5000~\mathrm{K}$, and find opacities in the range of $0.1 - 30~\mathrm{cm^2~g^{-1}}$, which increase for lower $Y_e$. One example of many light curve inferences is \cite{villar17}, who fit the multi-band light curves of GW170817 using a three-component ejecta described by $\kappa = 0.5,~3,~10~\mathrm{cm^{2}~g^{-1}}$ for blue (low opacity, lanthanide-poor), purple (moderate opacity), and red (high opacity, lanthanide-rich) ejecta components, respectively. By assuming these fixed opacities, they infer the ejecta masses, velocities, and temperatures of each component. However, in reality, the opacity may be strongly time-, velocity-, and wavelength-dependent (\eg, \citealt{kasen13, tanaka20, fontes20, kato24, deprince25}), as evidenced by the color evolution in the GW170817 kilonova.

In \cite{vieira23} and \cite{vieira24}, we introduce the Spectroscopic $r$-Process Abundance Retrieval for Kilonovae (\SPARK) software, which can be used to directly infer key kilonova ejecta properties from spectra, including $r$-process abundance patterns. \SPARK~fits kilonova spectra using the Monte Carlo radiative transfer code~\TARDIS~(\citealt{kerzendorf14}) through a Bayesian approach, accelerated by the Bayesian approximate posterior estimation scheme (\citealt{kandasamy17}) implemented in \approxposterior~(\citealt{fleming18}). \TARDIS~takes grids of elemental abundance patterns from nuclear reaction network calculations as an input, enabling us to retrieve the element-by-element abundances of $r$-process elements synthesized in the kilonova ejecta using \SPARK. In \cite{vieira23} and \cite{vieira24}, we model the spectrum of the GW170817 kilonova from 1.4 to 3.4 days, and infer the $r$-process abundance pattern at these epochs. Our modeling infers the presence of a single blue ejecta component at 1.4 and 2.4 days, while the 3.4 day spectrum requires a multi-component ejecta model containing an additional redder, lanthanide-rich component.

The abundance patterns inferred with \SPARK~originate from parametric nuclear network calculations (\citealt{wanajo18}), which also enable us to compute the radioactive heating rate $\dot{q}(t)$ given some ejecta composition. Furthermore, given the composition of the ejecta and other inferred properties such as temperature, density, and velocity, we can directly compute wavelength-, time-, and velocity-dependent opacities $\kappa(\lambda, t, v)$ in the ejecta, which can then be used to model the light curves.

In this work, we bridge spectral and light curve analyses of the GW170817 kilonova, and model both datasets. We infer the radioactive heating rates and opacities directly from the spectra of the GW170817 kilonova at 1.4, 2.4, 3.4, and 4.4 days post-merger using \SPARK. We use these inferred heating rates and opacities (rather than assuming a prescription for the heating rate or estimated gray opacity) to model the multi-band light curve evolution of the kilonova. Our spectral inference yields the detailed composition of the ejecta, while using a light curve model informed by our inferred heating rates and opacities allows us to infer the total mass of the ejecta. A schematic diagram illustrating our spectral and light curve modeling procedure is shown in Figure~\ref{fig:schematic_lightcurves}. Our approach thus demonstrates how leveraging both spectral and time series data together can provide additional insights which are not obtained from either approach alone.

\section{Methods}\label{sec:methods}

\subsection{Spectral Inference with SPARK}\label{ssc:spectralinference}

\begin{figure}
    \centering
    \includegraphics[width=0.46\textwidth]{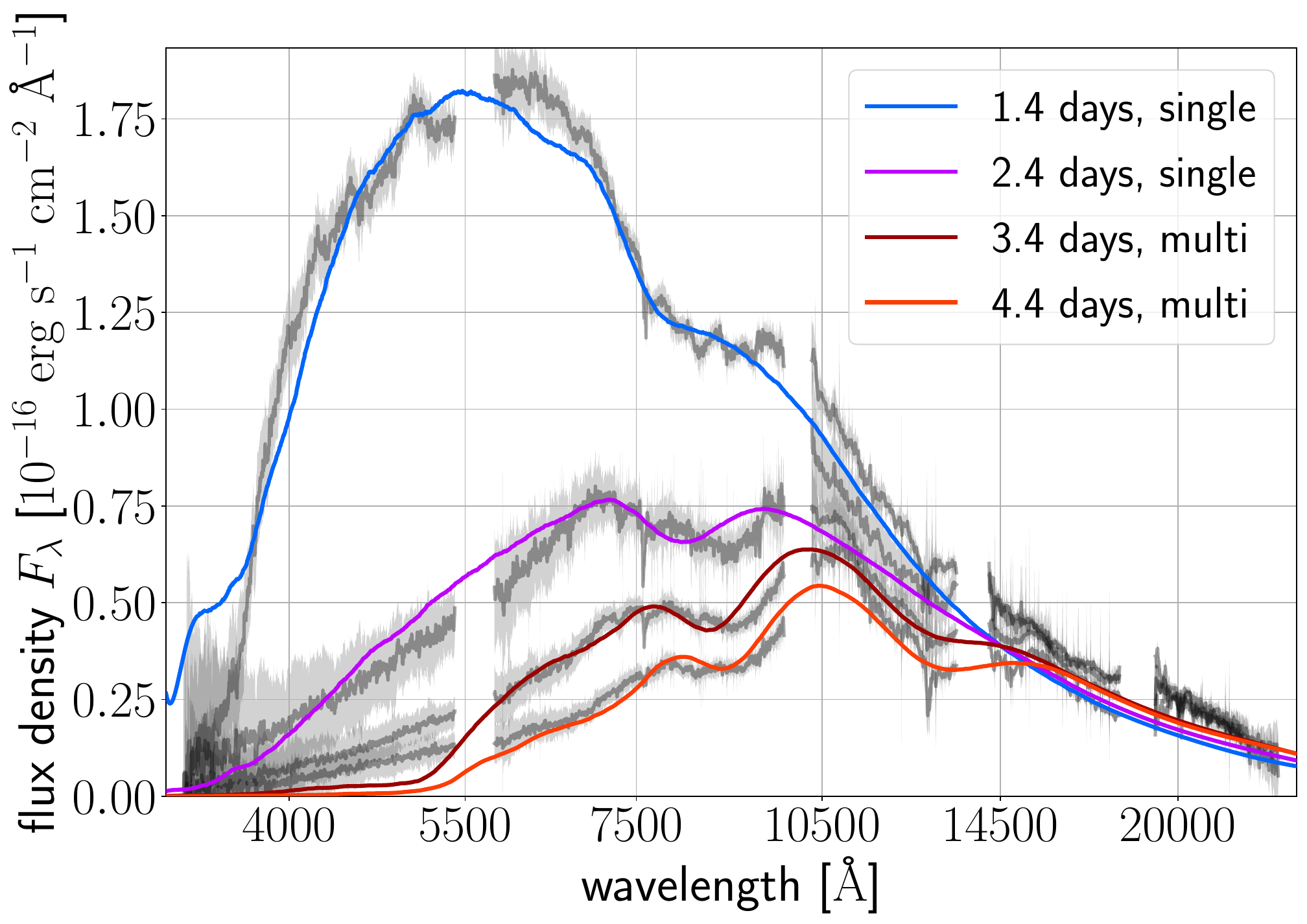}
    \caption{\textbf{Compilation of the best-fit synthetic spectra obtained with \SPARK, obtained by comparing radiative transfer simulations to the observed spectra of the GW170817 kilonova at 1.4, 2.4, 3.4, and 4.4 days post-merger.} The 1.4 and 2.4 day spectra are best-described by a single-component, bluer ejecta. At 3.4 and 4.4 days, an additional red ejecta component emerges, and thus the spectra are best-described with a two-component model. The corresponding best-fit abundance patterns are shown in Figure~\ref{fig:bestfit-abundances}.}
    \label{fig:bestfit-spectra}
\end{figure}

\begin{figure}
    \centering
    \includegraphics[width=0.48\textwidth]{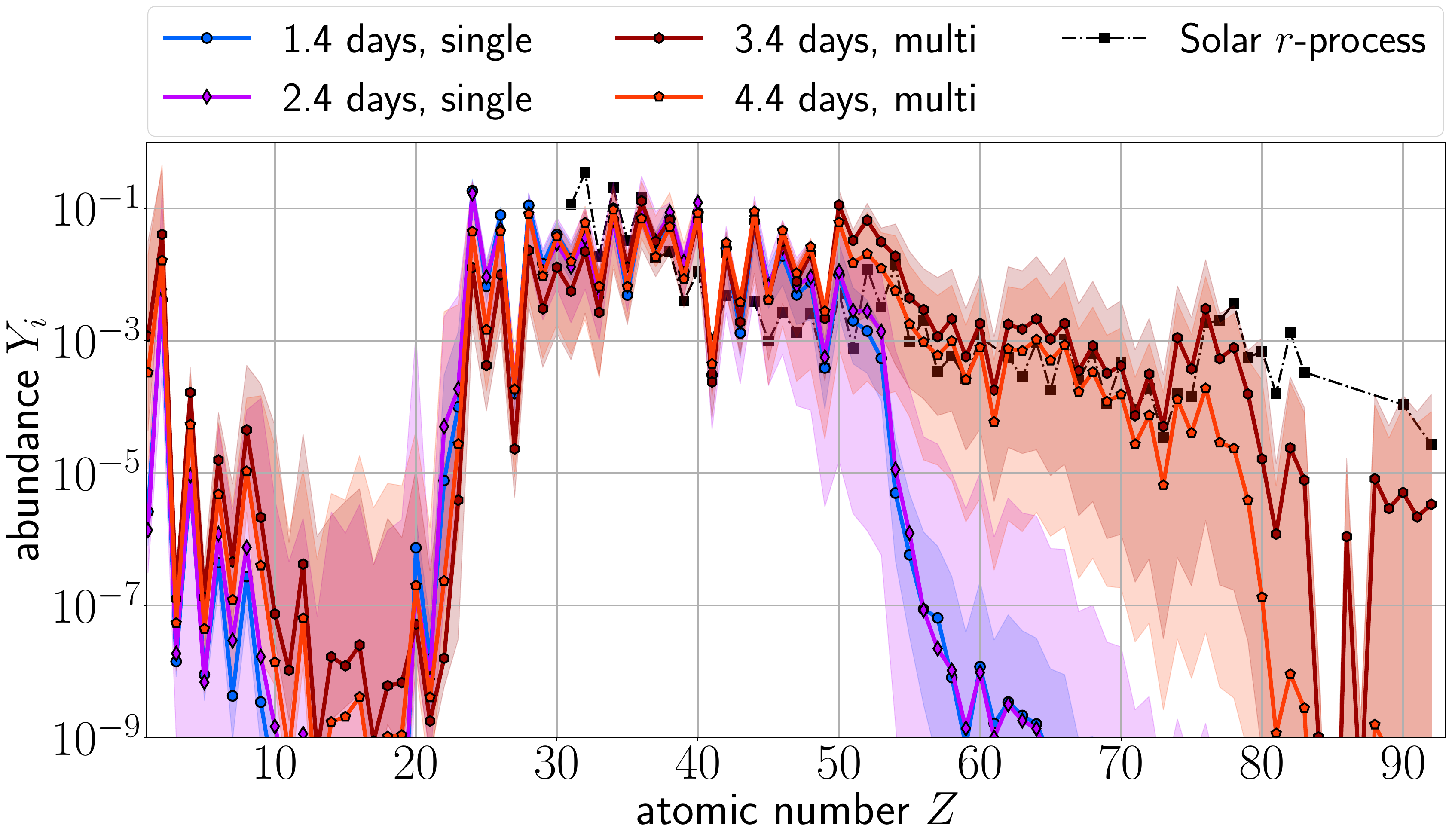}
    \caption{\textbf{Abundance patterns of the ejecta at 1.4, 2.4, 3.4, and 4.4 days, inferred from the spectra with \SPARK.} The abundances at 1.4 and 2.4 days are produced by a single blue component, while the ejecta at 3.4 and 4.4 days is multi-component, containing an additional red component that is richer in heavier elements. The corresponding best-fit spectra are shown in Figure~\ref{fig:bestfit-spectra}. We also show the Solar $r$-process abundance pattern for comparison, computed using the abundances from \cite{lodders09} with the $s$-process residual subtraction of \cite{bisterzo14}.}
    \label{fig:bestfit-abundances}
\end{figure}

In \cite{vieira23} (hereafter \vieiratwothree) and \cite{vieira24} (hereafter \vieiratwofour), we introduce \SPARK, our tool for fitting the spectra of kilonovae to infer the element-by-element abundances and other key properties of the kilonova ejecta. \SPARK~employs the 1D Monte Carlo radiative transfer code \TARDIS~(\citealt{kerzendorf14}) to synthesize spectra at a single point in time\footnote{We use \TARDIS~release \texttt{2022.06.05}, including full special relativity (\citealt{vogl19}) across \vieiratwothree, \vieiratwofour, and this work for consistency.}. The use of a spherically symmetric and non time-dependent code is required at present for computational feasibility, even with the accelerated inference facilitated by \approxposterior. That said, \SPARK~is written in a modular way such that one could swap \TARDIS~out for  another code with higher dimensionality. \TARDIS~assumes homologous expansion in the ejecta, which is accurate as of $\sim$$10^2 - 10^3$~s post-merger (\citealt{rosswog14}). \TARDIS~also relies on the Sobolev approximation, which assumes that (1) the thermal velocity is smaller than the velocity scale over which the ejecta varies and (2) the wavelength spacing of lines is larger than the thermal widths of the lines. The first condition is easily satisfied as the ejecta expands at velocities $\sim$$0.05 - 0.5c$~compared to thermal velocities of $\sim$$\mathrm{km~s^{-1}}$ (\citealt{kasen13}). \cite{tanaka20}~demonstrate that the second condition holds even for a suite of systematically computed theoretical lines for the lanthanides, which are the most numerous and most densely clustered in kilonova spectra.

In \vieiratwothree~and~\vieiratwofour, we fit the observed spectra of the GW170817 kilonova (\citealt{pian17, smartt17}) at 1.4, 2.4, and 3.4 days post-merger; here we also include a new fit to the spectrum at 4.4 days. Figure~\ref{fig:bestfit-spectra} shows our best-fit spectra at each of these epochs. Tables~\ref{tab:bestfit_single}~and~\ref{tab:bestfit_multi} present our best-fit parameters for the inference at each epoch, including the luminosity at the ejecta outer boundary $L_{\mathrm{outer}}$, density power law normalization $\rho_0 = \rho(t = t_0 = 1.5~\mathrm{days}, v=v_0=0.1c)$, inner and outer boundary velocities $v_{\mathrm{inner}}$~and~$v_{\mathrm{outer}}$, and abundance-setting parameters: electron fraction $Y_e,~\mathrm{expansion~velocity}~v_{\mathrm{exp}}$, and specific entropy $s$. For multi-component fits at 3.4 and 4.4 days, we infer distinct velocities, $Y_e,~v_{\mathrm{exp}},~\mathrm{and}~s$~for each of the two ejecta components. Uncertainties on inferred quantities (represented by bands or error bars throughout this work) are given by the 2.5\% and 97.5\% percentiles of their posteriors, respectively, unless stated otherwise. Importantly, these parameters are inferred given an inner boundary in the simulation that does not correspond to the photosphere, as detailed below. 

The ejecta in our single-component fits span a single shell with a unique velocity, density, temperature, and other plasma conditions, while the multi-component fits are described by 10 radial shells and thus a radial profile in these quantities. At initialization of a \TARDIS~simulation, the radiation temperature profile $T_{\mathrm{rad}}(v)$ is estimated given the inner boundary temperature $T_{\mathrm{inner}}$, which is itself initialized based on the user-input outer boundary luminosity $L_{\mathrm{outer}}$:

\begin{equation}
    T_{\mathrm{inner}} = \Big( \frac{L_{\mathrm{outer}}}{4 \pi \sigma (v_{\mathrm{inner}} t)^2} \Big)^{1/4} ,
\end{equation}

\noindent where $\sigma$~is the Stefan-Boltzmann constant. This temperature defines the peak wavelength of a blackbody by Wien's displacement law $\lambda_{\mathrm{peak,inner}} = b / T_{\mathrm{inner}}$, where $b = 2.898\times 10^7$~\AA~K. The temperature $T_{\mathrm{rad},i}$ in the $i$th shell is then related to a Doppler-shifted wavelength:

\begin{equation}
    \lambda_{\mathrm{peak},i}  = \lambda_{\mathrm{peak,inner}} (1 + \frac{v_{\mathrm{mid},i} - v_{\mathrm{inner}}}{c}) ,
\end{equation}

\begin{equation}
    T_{\mathrm{rad},i} = \frac{b}{\lambda_{\mathrm{peak},i}} = \frac{T_{\mathrm{inner}}}{1 + \frac{v_{\mathrm{mid},i} - v_{\mathrm{inner}}}{c}} ,
\end{equation}

\noindent where $v_{\mathrm{mid},i}$~is the velocity at the center of the $i$th shell. Finally, the dilution factor in each shell is initialized as the geometric dilution factor, \ie, $W(v) = \frac{1}{2}[ 1- (1 - (v_{\mathrm{inner}}/v)^2)^{1/2}]$~such that $W(v = v_{\mathrm{inner}}) = 0.5$. With this initial guess for $T_{\mathrm{inner}}$, $T_{\mathrm{rad}}(v)$, and $W(v)$, the plasma and radiation field are then iterated. The emitted luminosity is compared to the user-requested~$L_{\mathrm{outer}}$, and $T_{\mathrm{inner}}$ (and thus the inner boundary luminosity) is increased or decreased. The dilution factors $W$~and radiation temperature $T_{\mathrm{rad}}$~in each radial shell are iteratively estimated using Monte Carlo estimators, as described in \cite{kerzendorf14}. 

During this iteration, the dilution factor $W$~at the inner boundary may stray from its initial value of 0.5. Indeed, in the single shell for our single-component fits, and in the innermost shell for our multi-component fits, we find $W = 0.33^{+0.03}_{-0.03},~0.22^{+0.08}_{-0.05},~0.28^{+0.02}_{-0.02},~\mathrm{and}~0.28^{+0.02}_{-0.02}$ at 1.4, 2.4, 3.4, and 4.4 days, respectively. With this best-fit $W_ {\mathrm{inner}} \sim 0.3$, the inner boundary $v_{\mathrm{inner}}$~does not correspond to the continuum photosphere. This discrepancy challenges the assumption of \TARDIS~that the emission from the inner boundary is sampled from a blackbody distribution. However, with \SPARK,~we fit the observed sum of both a spectral continuum and absorption/emission features. The spectral continuum constrains quantities such as the temperature of an underlying blackbody, while the depth (height) of absorption (emission) features constrains the abundances of relevant species, which are in turn set by the aforementioned $Y_e,~v_{\mathrm{exp}}~\mathrm{and}~s$. The continuum and spectral features also both depend on the plasma conditions. We test the effect of decreasing $v_{\mathrm{inner}}$ with respect to our best-fit values, to attempt to approach the photosphere, observing changes in the continuum and spectral features. This lowering $v_{\mathrm{inner}}$~worsens the fit by increasing the optical depth of prominent lines, the depth of absorption features, and the height of emission features, due to the additional absorbing/emitting material included. Decreasing $v_{\mathrm{inner}}$~also does not raise $W_{\mathrm{inner}}$ beyond $\sim$$0.35$, down to $v_{\mathrm{inner}} = 0.1c$. Thus, while the best fit $v_{\mathrm{inner}}$~does not correspond to the continuum photosphere, the ejecta between $v_{\mathrm{inner}}$~and~$v_{\mathrm{outer}}$~does completely capture the line-forming region of the ejecta, given our model and inputs. Incorporating multi-dimensional effects, a more complete line list (as discussed later in this section), and more complex radial stratification in composition could reconcile these issues. In particular, we employ a composition which is the sum of two radially uniform components such that decreasing $v_{\mathrm{inner}}$~entails the presence of species which produce prominent spectral features at higher optical depths, leading to overestimated absorption/emission. Future work could employ more complex radially stratified compositions which enable the abundances of particular elements to decrease at smaller radii. This stratification could enable the best-fit $v_{\mathrm{inner}}$~to tend to the photosphere and $W_{\mathrm{inner}} \sim 0.5$ without worsening the fit. Alternatively, to resolve these issues of a lack of complete self-consistency, one could employ radiative transfer calculations which do not assume such an inner boundary and solve the plasma conditions in the entire ejecta, generating multi-band light curves and spectra simultaneously (\eg, \citealt{brethauer24, collins24, rahmouni25}). Such approaches to radiative transfer are however more computationally expensive, impeding the ability to perform statistically rigorous fitting and infer parameters of the kilonova's ejecta.

\begin{deluxetable}{ccc}[!t]
\centering
\tablecaption{Best-fit parameters for single-component fits to the GW170817 kilonova at 1.4 and 2.4 days. $v_{\mathrm{outer}}$ is fixed to $0.35c$ in the 1.4-day fit. Heating rates $\dot{q}$ are derived parameters, \ie, not varied in fitting the spectra.}
\tablehead{parameter & 1.4 days & 2.4 days}
\startdata\tablewidth{1.0\textwidth}
 \vspace{2pt}
$\log_{10}(\frac{L_\mathrm{outer}}{L_{\odot}})$ & $7.782^{+0.013}_{-0.014}$ & $7.594^{+0.040}_{-0.061}$ \\ 
$\log_{10}(\frac{\rho_0}{\mathrm{g~cm^{-3}}})$ & $-15.016^{+0.320}_{-0.316}$ & $-15.443^{+0.742}_{-0.463}$ \\ 
$v_{\mathrm{inner}}/c$ & $0.313^{+0.013}_{-0.014}$ & $0.249^{+0.017}_{-0.032}$ \\
$v_{\mathrm{outer}}/c$ & $0.35$ & $0.342^{+0.047}_{-0.050}$ \\
$v_{\mathrm{exp}}/c$ & $0.240^{+0.055}_{-0.082}$ & $0.172^{+0.107}_{-0.101}$ \\
$Y_e$ & $0.311^{+0.013}_{-0.011}$ & $0.306^{+0.055}_{-0.204}$  \\
$s~[k_{\mathrm{B}}/\mathrm{nuc}]$ & $13.6^{+4.1}_{-3.0}$ & $17.6^{+7.1}_{-6.3}$ \\ \hline
$\dot{q}~[10^{10}~\mathrm{erg~s^{-1}~g^{-1}}]$ & ${0.940}^{+0.128}_{-0.083}$ & ${0.484}^{+0.226}_{-0.347}$ \\
\enddata
\end{deluxetable}\label{tab:bestfit_single}

\begin{deluxetable}{ccc}[!t]
\centering
\tablecaption{Best-fit parameters for two-component fits to the GW170817 kilonova at 3.4 and 4.4 days. The radioactive heating rates $\dot{q}_{1}$, $\dot{q}_{2}$, and total $\dot{q}_{\mathrm{tot}}$ are derived parameters.}
\tablehead{parameter & 3.4 days & 4.4 days}
\startdata\tablewidth{1.0\textwidth}
 \vspace{2pt}
$\log_{10}(\frac{L_\mathrm{outer}}{L_{\odot}})$ & $7.605^{+0.049}_{-0.040}$ & $7.527^{+0.029}_{-0.034}$\\ 
$\log_{10}(\frac{\rho_0}{\mathrm{g~cm^{-3}}})$ & $-14.505^{+0.323}_{-0.372}$ & $-14.206^{+0.189}_{-0.357}$\\ \hline
$v_{\mathrm{inner,1}}/c$ & $0.213^{+0.056}_{-0.035}$ & $0.184^{+0.046}_{-0.022}$ \\
$v_{\mathrm{outer,1}}/c$ & $0.344^{+0.035}_{-0.038}$ & $0.337^{+0.041}_{-0.045}$ \\
$v_{\mathrm{exp,1}}/c$ & $0.198^{+0.092}_{-0.102}$ & $0.158^{+0.078}_{-0.078}$ \\
$Y_{e,1}$ & $0.228^{+0.073}_{-0.088}$ & $0.209^{+0.116}_{-0.113}$ \\
$s_{1}~[k_{\mathrm{B}}/\mathrm{nuc}]$ & $14.9^{+8.0}_{-4.5}$ & $18.6^{+9.5}_{-7.3}$ \\ \hline
$v_{\mathrm{inner,2}}/c$ & $0.232^{+0.038}_{-0.027}$ & $0.172^{+0.047}_{-0.024}$ \\
$v_{\mathrm{outer,2}}/c$ & $0.334^{+0.037}_{-0.022}$ & $0.329^{+0.056}_{-0.060}$ \\
$v_{\mathrm{exp,2}}/c$ & $0.134^{+0.091}_{-0.069}$ & $0.235^{+0.060}_{-0.107}$\\
$Y_{e,2}$ & $0.161^{+0.149}_{-0.104}$ & $0.280^{+0.098}_{-0.162}$ \\
$s_{2}~[k_{\mathrm{B}}/\mathrm{nuc}]$ & $21.5^{+8.3}_{-9.5}$ & $16.2^{+7.3}_{-5.4}$ \\ \hline
$\dot{q}_1~[10^{10}~\mathrm{erg~s^{-1}~g^{-1}}]$ & ${0.427}^{+0.116}_{-0.116}$ & ${0.419}^{+0.065}_{-0.260}$ \\
$\dot{q}_2~[10^{10}~\mathrm{erg~s^{-1}~g^{-1}}]$ & ${0.322}^{+0.341}_{-0.071}$ & ${0.237}^{+0.142}_{-0.207}$ \\ \hline
$\dot{q}_{\mathrm{tot}}~[10^{10}~\mathrm{erg~s^{-1}~g^{-1}}]$ & ${0.381}^{+0.173}_{-0.067}$ & ${0.328}^{+0.031}_{-0.187}$ \\
\enddata
\end{deluxetable}\label{tab:bestfit_multi}

Our atomic physics assumptions and choices are consistent with \vieiratwothree~and~\vieiratwofour. We use the same line list, which contains $\sim$$1.5$~million empirical lines from the Vienna Atomic Line Database (VALD;  \citealt{ryabchikova15, pakhomov19}) and 19 additional lines for species of interest (neodymium ${}_{60}$\ion{Nd}{2}, \citealt{hasselquist16}~and cerium ${}_{58}$\ion{Ce}{2}, \citealt{cunha17}) from the APOGEE survey (\citealt{majewski17}). However, the incompleteness of these line lists is a systematic source of uncertainty in modeling kilonova spectra (\eg, \citealt{kasen17, watson19, gillanders22, gillanders24, sneppen24}). This incompleteness prevents us from capturing the complete bound-bound opacity. Substantial efforts have been made towards systematic theoretical atomic structure calculations to supplement the dearth of empirical lines (\eg, \citealt{tanaka20, fontes20, kato24, deprince25}), but these calculations yield non-negligible differences in energy levels, transition wavelengths, and transition probabilities. \cite{tanaka20} and \cite{kato24}~demonstrate that differing strategies, even given the same atomic structure code, yield opacities for the singly-ionized lanthanides which differ by as much as 50\%. \cite{brethauer24}~generate kilonova light curves and spectra for three different theoretical atomic data sets, and find order-of-magnitude differences in the lanthanide fraction of the ejecta and 25-40\% differences in the total ejecta mass. We thus opt to keep the same line lists as used in our prior works, despite the incomplete bound-bound opacity. We briefly discuss the impact of this incompleteness in Section~\ref{ssc:disco-opacities}; future work could quantify how the incompleteness of the line lists and differences between theoretical codes impact the plasma conditions in kilonovae in greater detail.

Our nuclear physics assumptions are also consistent with \vieiratwothree~and~\vieiratwofour. We use the nuclear reaction network calculations of \cite{wanajo18} as our source of abundance patterns, which parametrize the abundances as a function of electron fraction $Y_e$, expansion velocity $v_{\mathrm{exp}}$, and specific entropy $s$. Spectral modeling with \TARDIS~(\citealt{kerzendorf14}) takes in these abundance patterns as an input, enabling us to directly infer the abundances as a function of the $Y_e$, $v_{\mathrm{exp}}$, and $s$ parameters. Our best-fit abundance patterns at the four spectral epochs are shown in Figure~\ref{fig:bestfit-abundances}. 

The spectra at 1.4 and 2.4 days poster-merger are best described by a single-component, bluer (higher-$Y_e$) ejecta that contains a higher abundance of lighter $r$-process elements. At 3.4 days, and persisting at 4.4 days, an additional redder, lower-$Y_e$ ejecta component with an high abundance of heavier elements emerges. This redder component has a substantial abundance of lanthanides (atomic number $Z \in [57,71]$), characterized by high opacities in the ultraviolet and optical (\citealt{tanaka20, kato24}), leading to the redder color observed in the spectra. Notably, our inference does not suggest a strong radial gradient in the composition: the blue and red components nearly completely overlap radially at 3.4 and 4.4 days, although \TARDIS~is only one-dimensional and thus we cannot comment on the angular distribution. This near-complete overlap also means that the mass is roughly equiparitioned between the two components. 

The differing values in our inferred $\rho_0$~for different epochs do not imply evolution in this quantity, which is defined by the same $t_0 = 1.5~\mathrm{days}$~and~$v_0=0.1c$~at all epochs. Rather, it indicates evolution in our inferred value. Given the range of inferences across all epochs, we may conservatively state that $\log_{10}(\rho_0 / \mathrm{g~cm^{-3})}$~lies in the range $\sim$$-15.4$~to~$-14.2$. Indeed, we do not strongly constrain the density (or equivalently mass), motivating our light curve fits. Future work could fit multiple epochs simultaneously, tying together $\rho_0$ throughout all epochs, potentially obtaining a tighter constraint from the spectra. 

We refer the reader to \vieiratwofour~and other relevant works (\citealt{watson19,domoto22,gillanders22,sneppen24}) for in-depth discussions on the evolution of these parameters and their imprint on the emergent spectra, and \cite{ford24}~for a wider parameter space exploration of kilonova spectra with \TARDIS. We now turn our attention to the quantities of greatest interest for the light curves: the radioactive heating rates and opacities.

\subsection{Inferring Radioactive Heating Rates}\label{ssc:infer-heatingrates}

\begin{figure}[!ht]
    \centering
    \includegraphics[width=0.48\textwidth]{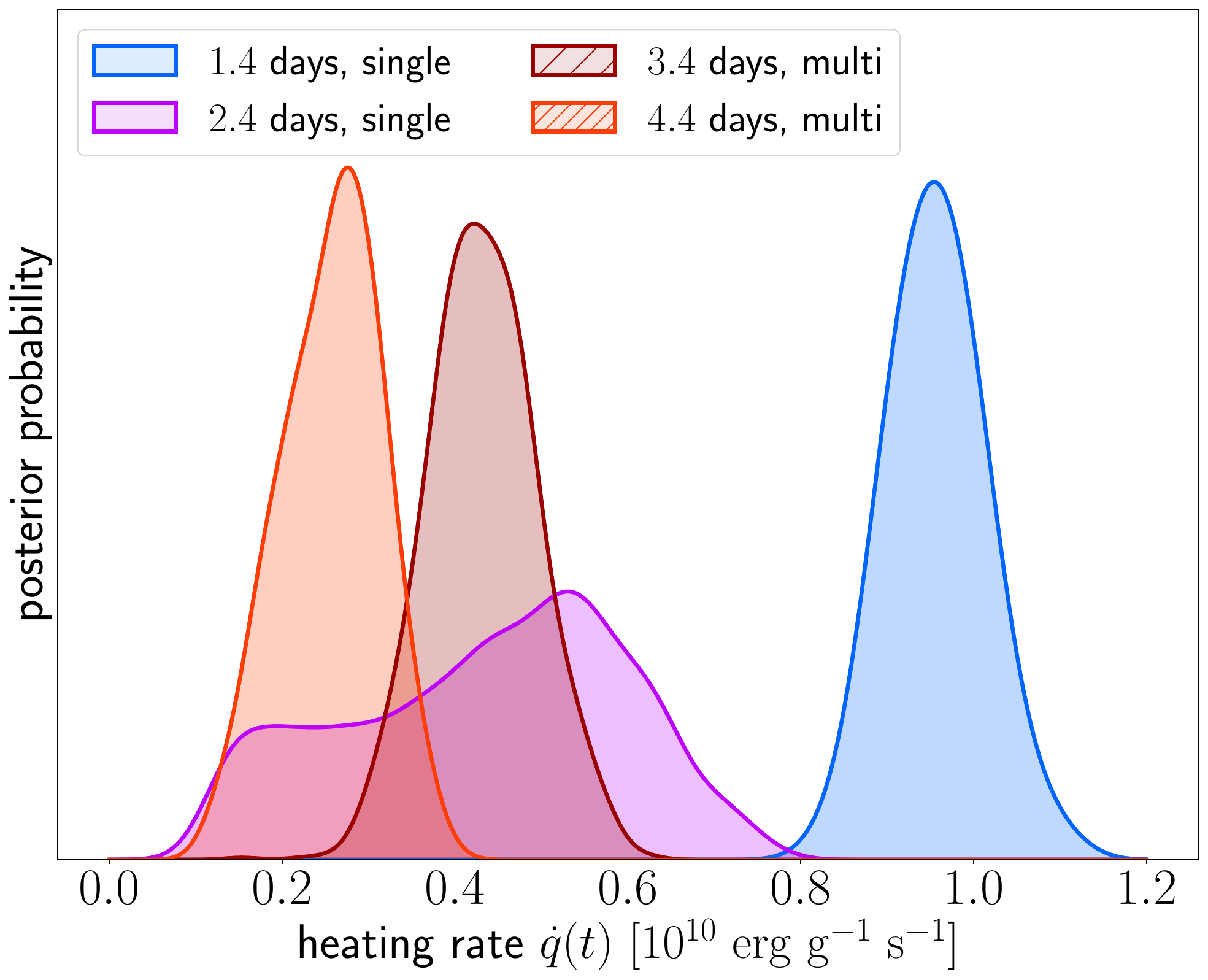}
    \caption{\textbf{Inferred posterior distributions of the radioactive heating rates of the ejecta at 1.4, 2.4, 3.4, and 4.4 days.} As with the abundances (Figure~\ref{fig:bestfit-abundances}), the heating rates are extracted from the nuclear network calculations of \cite{wanajo18} given the inferred $Y_e,~v_{\mathrm{exp}},~\mathrm{and}~s$ at each epoch. At 3.4 and 4.4 days, our best-fit models are multi-component, and each of the two components has its own composition and thus its own inferred heating rate; we report the mass-weighted sum of both components. Appendix~\ref{app:radioactive-components} includes posteriors for the heating rates of these distinct components.}
    \label{fig:hists-heatingrates}
\end{figure}

\begin{figure}[!ht]
    \centering
    \includegraphics[width=0.48\textwidth]{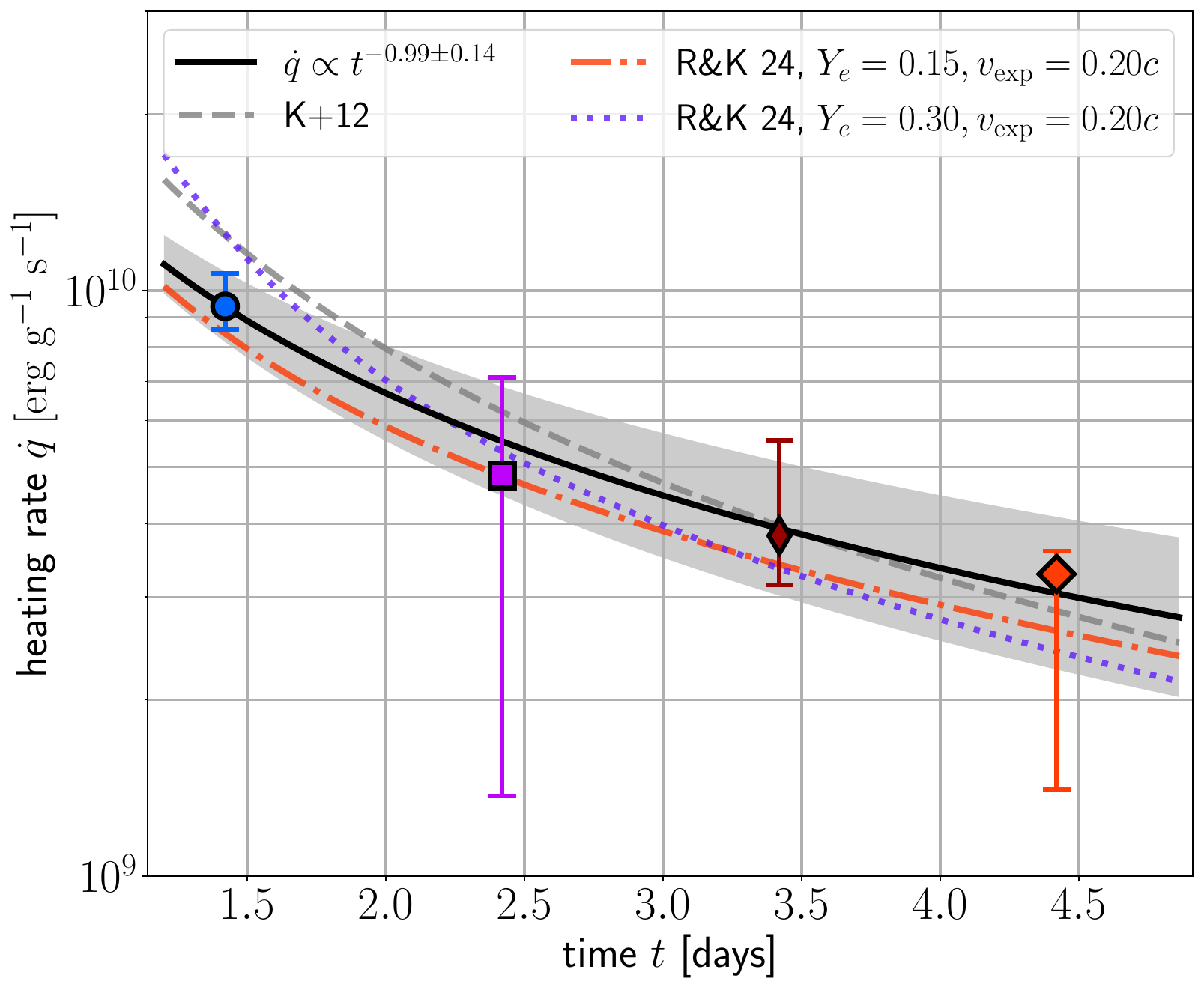}
    \caption{\textbf{Inferred total radioactive heating rates of the ejecta as a function of time.}  As with the abundances, the heating rates are extracted from the nuclear network calculations of \cite{wanajo18}, given the inferred $Y_e,~v_{\mathrm{exp}},~\mathrm{and}~s$ at each epoch. The total heating rates at 3.4 and 4.4 days are the mass-weighted total heating rate of the red and blue components at each respective epoch. Fitting the inferred heating rates, we find a power-law of $\dot{q}(t) = \dot{q_0} (t / {\mathrm{1~day}})^\alpha$ with $\dot{q_0} = 10^{10.12 \pm 0.04}~\mathrm{erg~g^{-1}~s^{-1}}$ and $\alpha = -0.99 \pm 0.14$. This fit is shown as solid line, and the gray shaded band reflects uncertainties in $\dot{q_0}$ and $\alpha$. For comparison, we include the heating rate fit of \cite{korobkin12} (K+12), which holds primarily for $Y_e \approx 0.05$, and the more broadly-applicable fit from \cite{rosswog24} (R\&K 24) for two fiducial sets of $(Y_e,~v_{\mathrm{exp}})$.} 
    \label{fig:points-heatingrates}
\end{figure}

\begin{figure}
    \centering
    \includegraphics[width=0.47\textwidth]{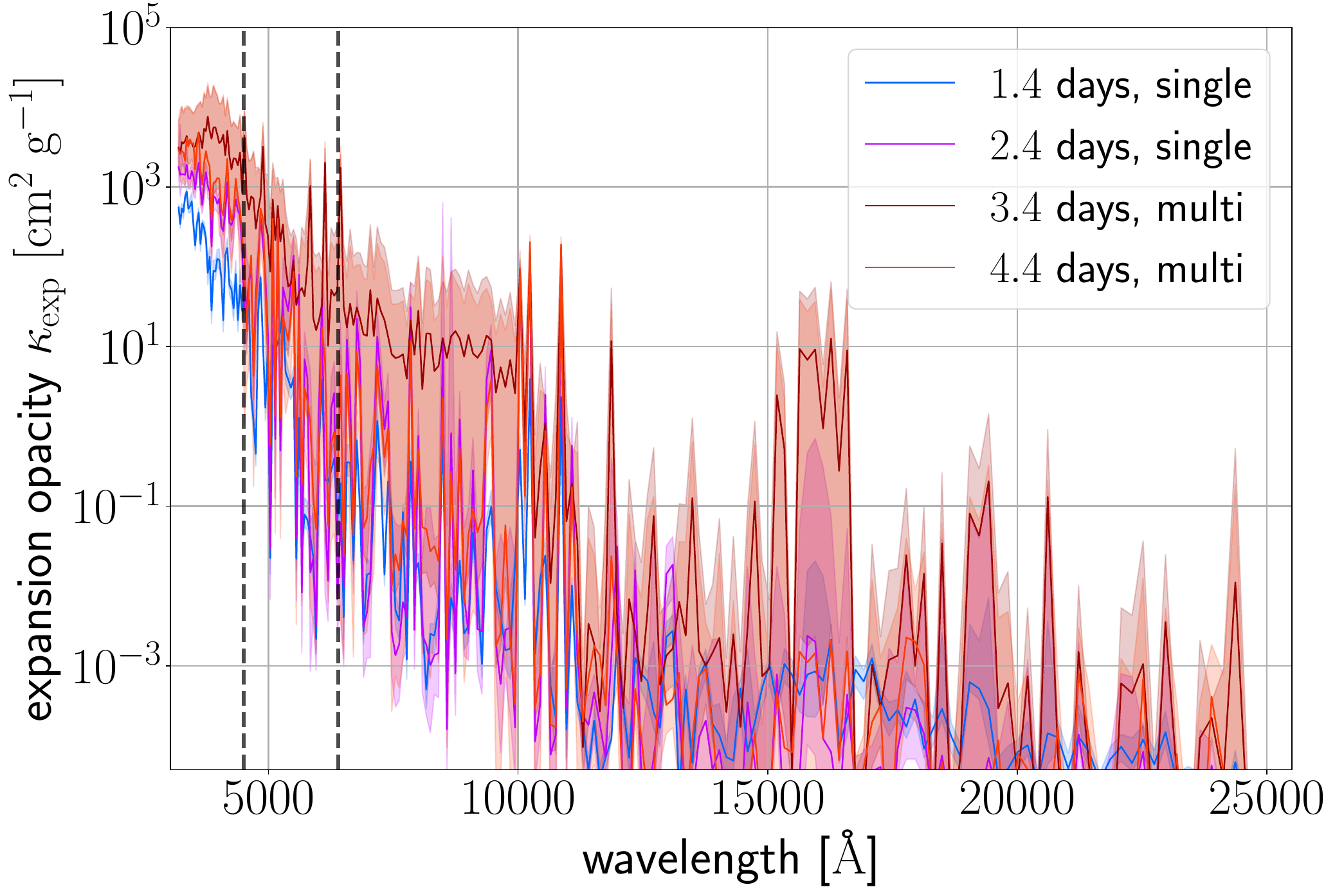}
    \caption{\textbf{Wavelength-dependent expansion opacities of the ejecta, inferred at 1.4, 2.4, 3.4, and 4.4 days.} Shaded bands represent the uncertainties on opacities based on the uncertainties on the composition at each epoch. Opacities are computed assuming a single shell of material, spanning some $v_{\mathrm{inner}}$ and $v_{\mathrm{outer}}$ as given by our \SPARK~fits, at a unique temperature and density. This $v_{\mathrm{inner}}$ lies above the photosphere. We adjust $\rho_0 = \rho(t = 1.5~\mathrm{days}, v=0.1c)$ in the power-law density profile such that there is $0.01~M_{\odot}$ of ejecta between $v = 0.1c$ and $0.4c$ at 1.5 days post-merger. We set the temperature to that obtained from spectral fitting (\citealt{sneppen24}): $5440,~3940,~3420,~\mathrm{and}~3330~\mathrm{K}$ at 1.4, 2.4, 3.4, and 4.4 days, respectively. Corresponding wavelength-integrated Planck mean opacities are shown as a function of velocity in Figure~\ref{fig:planck-opacities} and time in Figure~\ref{fig:planck-opacities-fitt}. We mark with dashed gray lines (at $4500$~\AA~and~$6400$~\AA) the relevant wavelength cutoffs for computing our Planck mean opacities (at 1.4, 2.4 days and 3.4, 4.4 days, respectively).}
    \label{fig:expansion-opacities}
\end{figure}

\begin{figure*}
    \centering
    \includegraphics[width=0.48\textwidth]{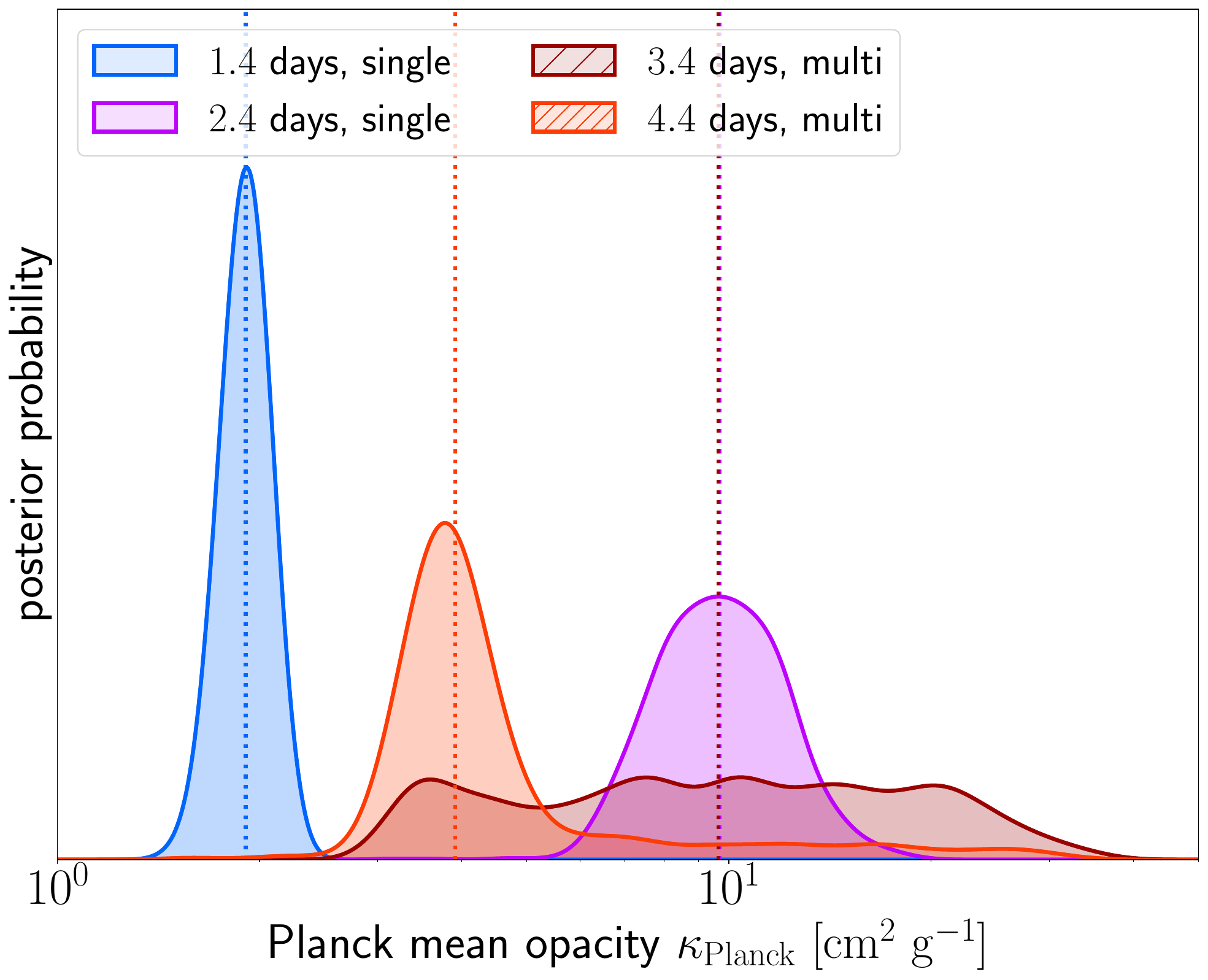}
    \includegraphics[width=0.47\textwidth]{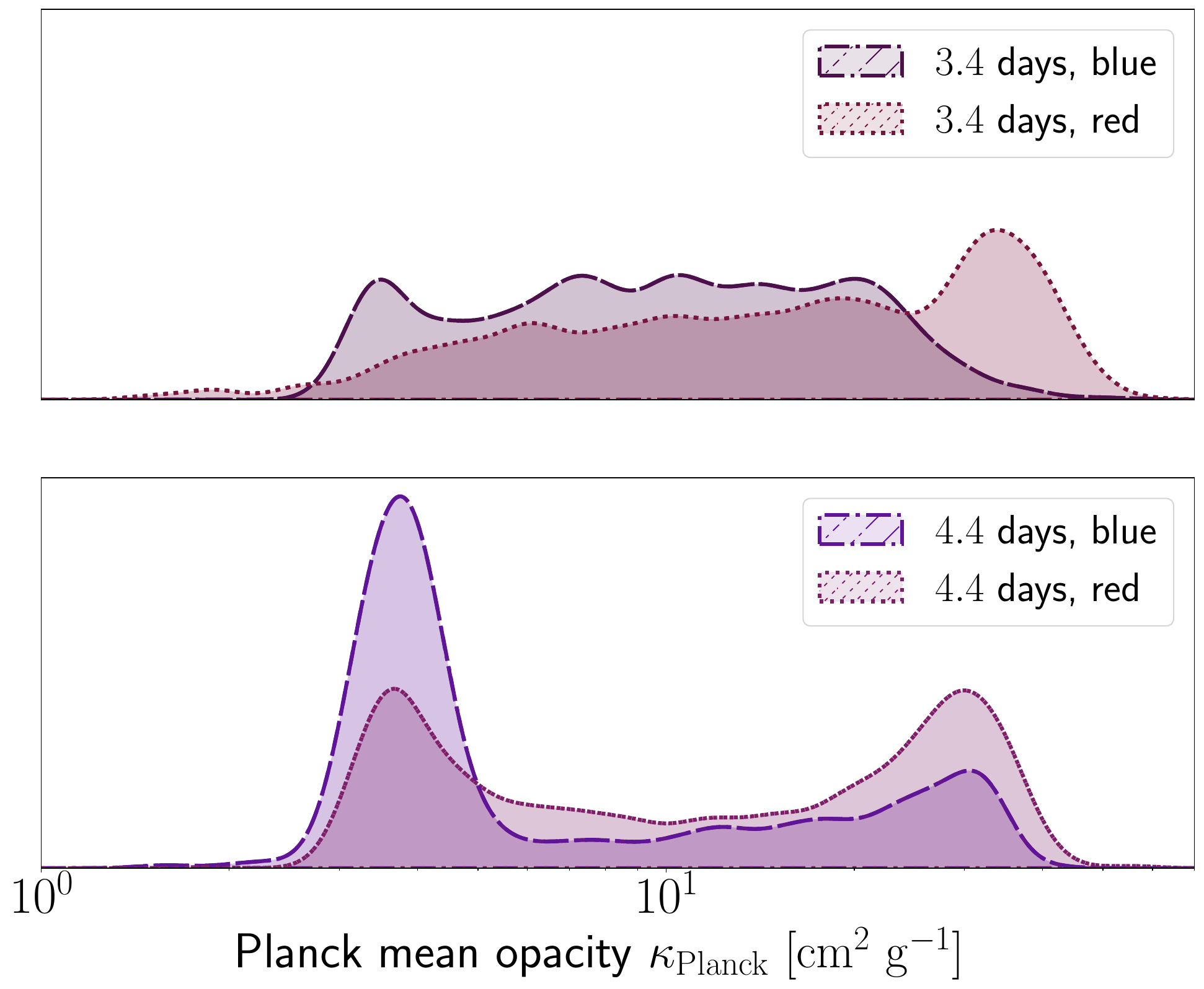}
    \caption{\textbf{Inferred posterior distributions of the Planck mean opacities at 1.4, 2.4, 3.4, and 4.4 days.} Given wavelength-dependent expansion opacities (Figure~\ref{fig:expansion-opacities}), integrating over wavelength weighted by the Planck function at some $T$~yields the Planck mean opacity. \textit{Left:} Total opacities at each epoch. Dotted vertical lines indicate the best-fit (posterior median) opacities which are input into the light curve model. These posterior medians are indistinguishable at 2.4 and 3.4 days. At 3.4 and 4.4 days, our best-fit models are multi-component, and each of the two components has its own composition. We report the opacity obtained by mass-weighting and mixing the two components \textit{a priori} and computing the opacity of this mixture of two components. This approach captures the physics of reprocessing, whereby one component may absorb the emission of the other and reemit it. \textit{Right}: Opacities of these individual components, with their own compositions, at 3.4 and 4.4 days.}
    \label{fig:hists-opacities}
\end{figure*}

\begin{figure}
    \centering
    \includegraphics[width=0.48\textwidth]{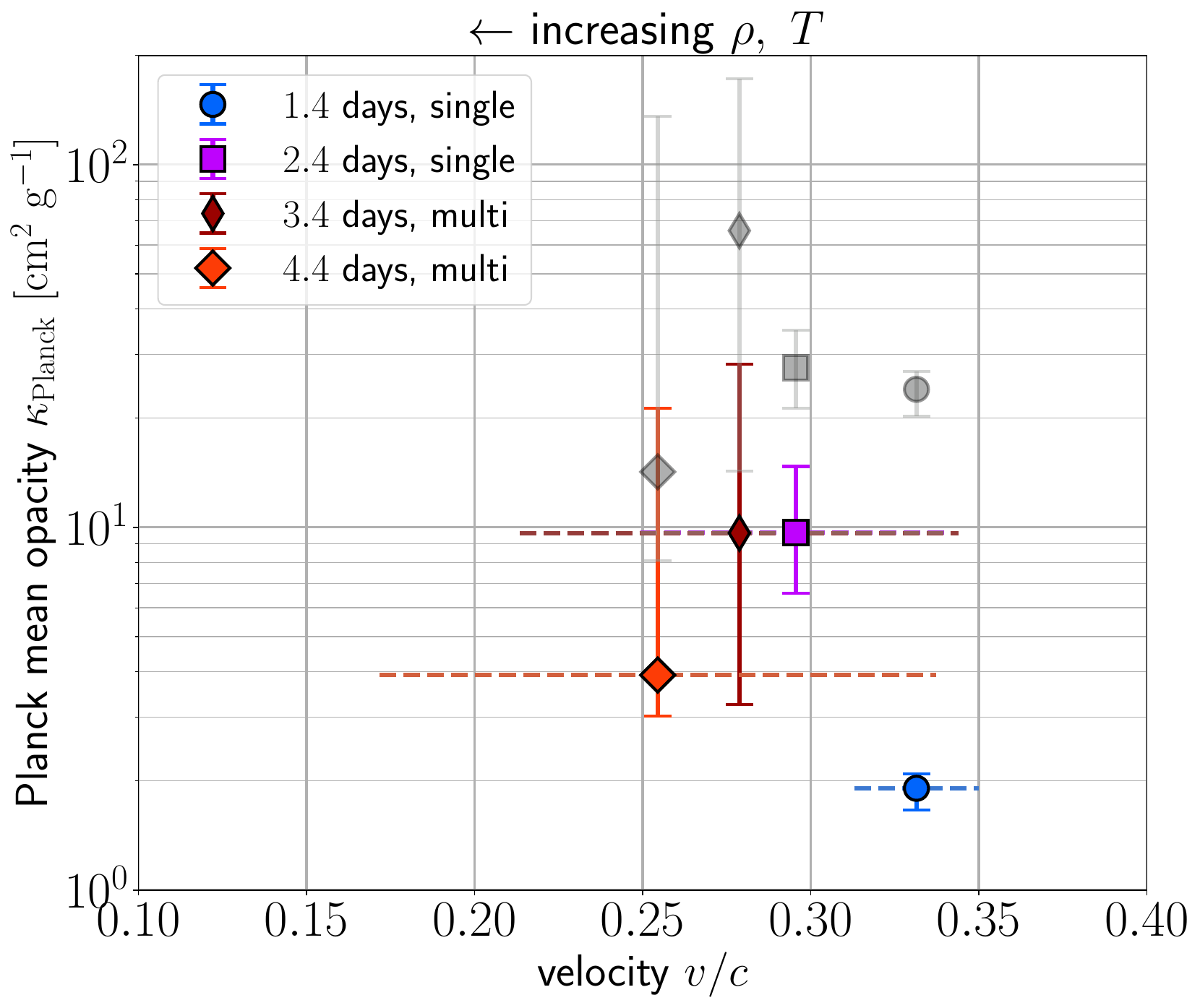}
    \caption{\textbf{Planck mean opacities, as a function of radial velocity, inferred at 1.4, 2.4, 3.4, and 4.4 days.}  At each epoch, the horizontal dashed line indicates the span of the ejecta (from $v_{\mathrm{inner}}$ and $v_{\mathrm{outer}}$) in the line-forming region as inferred from \SPARK. This $v_{\mathrm{inner}}$ lies above the photosphere. Gray markers show the opacities without imposing a $\leqslant4500$~\AA~cutoff at 1.4 and 2.4 days and $\leqslant6400$~\AA~cutoff at 3.4 and 4.4 days; solid colored markers show the (preferred) opacities with these cutoffs imposed. Opacities are dependent on velocity due to the radial temperature and density profiles: the expansion opacity goes as $1/\rho$ (where $\rho \propto v^{-3}$), reducing the opacities at higher densities. This strong density/velocity dependence leads to an opacity which mostly increases with $v$, except at 1.4 days where the higher temperature changes the ionization fractions of the ejecta and the weighting Planck function, resulting in a lower opacity.}
    \label{fig:planck-opacities}
\end{figure}

\begin{figure}
    \centering
    \includegraphics[width=0.48\textwidth]{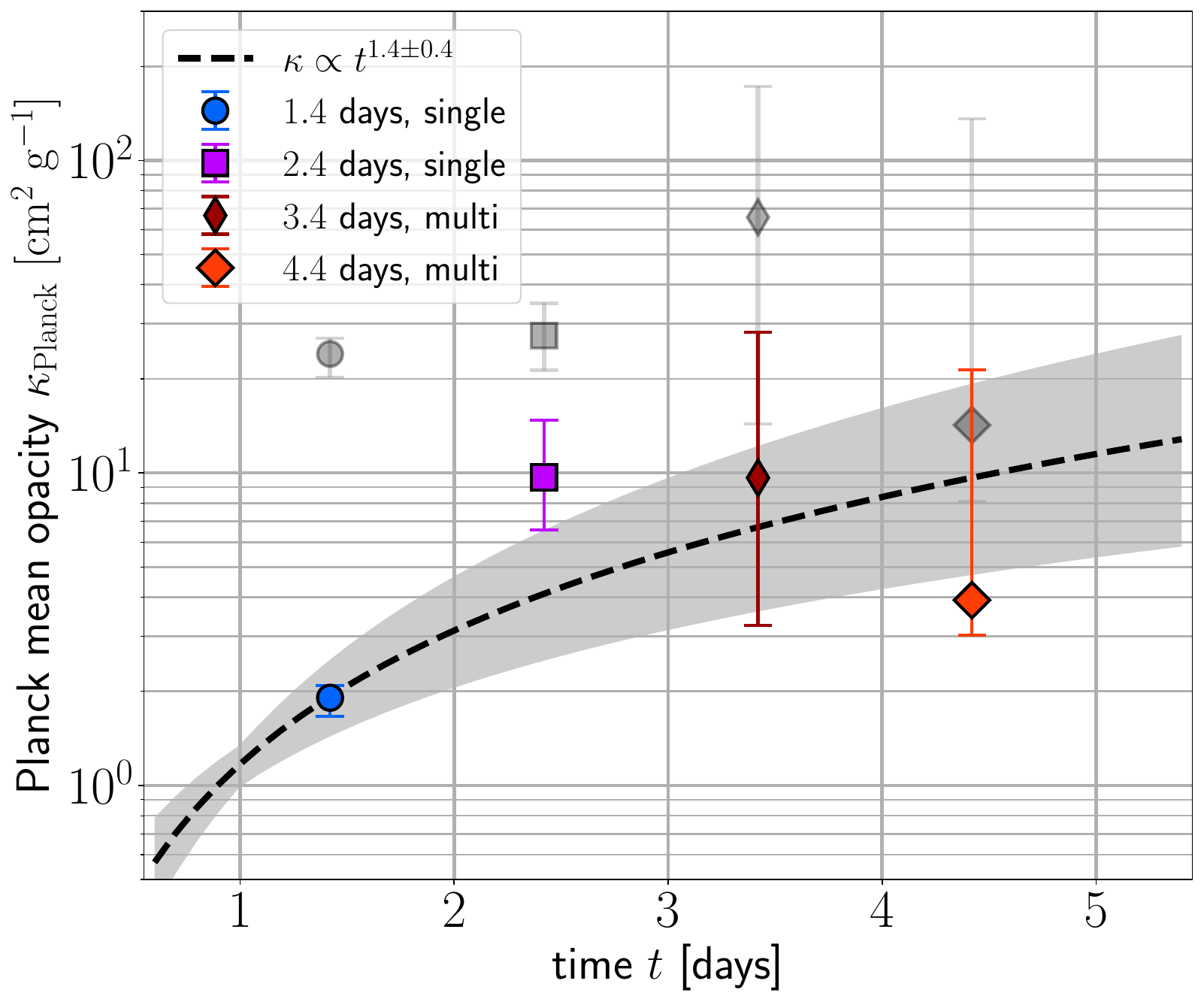}
    \caption{\textbf{Planck mean opacities, as a function of time, inferred at 1.4, 2.4, 3.4, and 4.4 days.} The opacities in time are well described by a power-law $\kappa(t) = \kappa_0 (t / ~\mathrm{1~day})^{\gamma}$ with $\kappa_0 = 1.17 \pm 0.18~\mathrm{cm^2~g^{-1}}$ and $\gamma = 1.4 \pm 0.4$, shown as a dashed line, with the gray translucent band reflecting uncertainties on $\kappa_0$ and $\gamma$. Gray markers show the opacities without imposing a $\leqslant4500$~\AA~cutoff at 1.4 and 2.4 days and $\leqslant6400$~\AA~cutoff at 3.4 and 4.4 days when computing Planck mean opacities; solid colored markers show the (preferred) opacities with these cutoffs imposed.}
    \label{fig:planck-opacities-fitt}
\end{figure}

We use the set of parametric trajectories from the nuclear network calculations of \cite{wanajo18} to obtain the composition and radioactive heating rates at given our inferred $Y_e,~v_{\mathrm{exp}},~\mathrm{and}~s$. Given a nuclear network calculation, tracer particles propagate through a plasma to trace the composition of the ejecta as a function of three parameters: $Y_e$, $v_{\mathrm{exp}}$, and $s$~in the plasma. Crucially, \cite{wanajo18}~compute the composition and heating rates as a function of time, from $10^{-8}~\mathrm{days}$~to~$\sim$$1~\mathrm{year}$~post-merger; we use this time dependence. Figure~\ref{fig:hists-heatingrates} shows our inferred posterior probability distributions of $\dot{q}(t)$ at 1.4, 2.4, 3.4, and 4.4 days post-merger. We emphasize that these should not be interpreted as distributions of $\dot{q}(t)$ in the ejecta material, but rather a probabilistic distribution of the best-fit parameters. At 3.4 days and 4.4 days, where we infer the presence of a new lower-$Y_e$ redder ejecta component in addition to the earlier higher-$Y_e$ bluer ejecta component, we infer separate heating rates for the blue and red components. Since our light curve model (Section \ref{ssc:lightcurve-model}) requires a single heating rate at each epoch, we compute a total heating rate for the two components at 3.4 and 4.4 days. This total heating rate is the mass-weighted sum of both ejecta components. Posterior distributions for the heating rates of each distinct component are provided in Appendix~\ref{app:radioactive-components}.

Figure~\ref{fig:points-heatingrates} shows our inferred total heating rate as a function of time. Given the roughly equal partition in mass between the red and blue ejecta components due to the lack of a radial composition gradient, the heating rates at 3.4 and 4.4 days post-merger are roughly the average of the individual components' heating rates at these epochs. We compare these inferred heating rates to the semi-analytic fit of \cite{korobkin12} (most appropriate for $Y_e \approx 0.05$) and the more broadly-applicable fits of \cite{rosswog24} in Figure~\ref{fig:points-heatingrates}. We show the \cite{rosswog24} heating rates for two fiducial scenarios: a bluer ejecta dominated by lighter $r$-process elements ($Y_e = 0.30,~v_{\mathrm{exp}} = 0.20c$), as we see at 1.4 and 2.4 days, and a redder ejecta ($Y_e = 0.15,~v_{\mathrm{exp}} = 0.20c$), as emerges at 3.4 and 4.4 days. Our \SPARK~inferred heating rates are largely consistent with these alternative descriptions, and follow the expected power-law behavior. Indeed, fitting a simple power-law $\dot{q_0} (t / {\mathrm{1~day}})^\alpha$ to our inferred heating rates, we find $\dot{q_0} = 10^{10.12 \pm 0.04}~\mathrm{erg~g^{-1}~s^{-1}}$ and $\alpha = -0.99 \pm 0.14$. This is somewhat less steep than the oft-used $\alpha = -1.3$. The most constraining point in the fit comes from our inference at 1.4 days. 

It is non-trivial that the heating rates inferred from the spectra, for ejecta $v \gtrsim 0.17c$, should also be applicable to the remainder of the ejecta. Indeed, most of the ejecta mass, which dominates the heating rate, is located at smaller velocities, including below the photosphere. That we are only modeling a fraction of the ejecta with our spectral model is especially clear given that our inner boundary $v_{\mathrm{inner}}$, with $W_{\mathrm{inner}} \sim 0.3 < 0.5$, lies above the photosphere. Feeding these inferred heating rates (and opacities, discussed in the next section) into a light curve model serves as a test of the extendability of our inferred quantities to the remainder of the ejecta, outside the line-forming region.

\subsection{Inferring Time and Wavelength-Dependent Opacities}\label{ssc:infer-opacities}

\subsubsection{Expansion Opacities}

Given a sample from our \SPARK~posteriors for $Y_e,~v_{\mathrm{exp}}$~and~$s$, we also have all the ingredients to extract an opacity. \TARDIS~uses the Sobolev approximation to describe the optical depth of bound-bound transitions in a homologously-expanding ejecta. For each sample in our posterior, we instantiate a plasma and perform radiative transfer under the assumptions of local thermodynamic equilibrium (LTE). Level populations and ionization are computed using the Saha-Boltzmann equations. This computation yields the Sobolev optical depth $\tau_{\mathrm{Sob}}$ for all relevant atomic transitions. We then compute the wavelength-dependent expansion opacity $\kappa_{\mathrm{exp}}(\lambda)$, which sums transitions in successive wavelength bins of width $\Delta\lambda$, weighted by $(1 - e^{-\tau_{\mathrm{Sob}}})$. 

Figure~\ref{fig:expansion-opacities} shows the expansion opacities inferred at 1.4, 2.4, 3.4, and 4.4 days, along with uncertainties. To obtain a reasonable opacity, we must use some total ejecta mass beyond what is contained in the line-forming region. We thus compute opacities for $0.01~M_{\odot}$ of ejecta between $0.1c$ and $0.4c$ at 1.5 days post-merger, in line with other opacity calculations (\eg, \citealt{tanaka20}). To set the temperatures for the opacity calculations, we first perform a blackbody fit to the entire spectral range at each epoch and find~$4800,~3400,~2900,~\mathrm{and}~2700~\mathrm{K}$ at 1.4, 2.4, 3.4, and 4.4 days, respectively. However, \cite{sneppen24}~point out that such a simple fit to the entire spectral range does not account for deviations from a blackbody due to emission/absorption, which can bias this measurement to colder temperatures at early times. Fitting more carefully with additional parameters to capture the absorption/emission, \cite{sneppen24}~find temperatures of $5440,~3940,~3420,~\mathrm{and}~3330~\mathrm{K}$~at 1.4, 2.4, 3.4, and 4.4 days, respectively. We thus use these more carefully-measured temperatures in our calculations and throughout all of our analyses. 

We compute the expansion opacities in a single shell of material spanning the inferred $v_{\mathrm{inner}}$ and $v_{\mathrm{outer}}$, \ie, at a single density and temperature. In all calculations, we compute the opacities for a single composition. In the case of the two-component ejecta, we compute the opacity for the mass-weighted and summed mixture of the red and blue components. This approach is distinct from the more common practice of computing the opacities for different components separately and/or computing emission from distinct components and summing those emissions. Our approach here thus captures the important physics of reprocessing, whereby one component can absorb the emission of the other and reemit it, leading to a diversity of kilonovae (\eg, \citealt{kawaguchi20}). This approach yields a single wavelength-dependent opacity at each epoch. 

The expansion opacities in Figure~\ref{fig:expansion-opacities} display the strong wavelength dependence that is expected. At 3.4 and 4.4 days, the opacities are overall higher due to the larger abundance of open $d$-shell ($Z \in [40,~58]$~and $[72,~80]$) and $f$-shell elements, especially the lanthanides. The opacities are highest at the shortest wavelengths $\lesssim$$10,000$ \AA, due to the presence of these lanthanides with high opacities at these wavelengths. There is an additional peak in opacity at $\sim$$15,000$~\AA, which we attribute to \ion{Ce}{2}~in \vieiratwofour.

\subsubsection{Time-dependent Gray Opacities}

With these expansion opacities in hand, we require wavelength-independent (`gray') opacities that can be ingested by our simple light curve model. As we will elaborate in Section \ref{ssc:lightcurve-model}, we opt for a simple light curve model with gray opacities, the fits of which can be compared to the fits obtained from other simple models commonplace in light curve modeling (\eg, \citealt{villar17}). To obtain gray opacities, we compute the Planck mean opacity. This Planck mean opacity is the mean of the expansion opacity over wavelength, weighted by the Planck function at some temperature $T$. Importantly, in \vieiratwothree~and \vieiratwofour, we find that our spectral fits (see Figure~\ref{fig:bestfit-spectra}) perform poorly at $\lesssim$$4500$~\AA~at 1.4 and 2.4 days. This poor performance can be attributed in part to overestimated absorption from yttrium (${}_{39}$Y). At 3.4 and 4.4 days, we similarly see poor performance, where the fit is biased towards a blackbody unless short wavelengths are excluded. This is resolved by excluding data $\leqslant 6400$~\AA~\textit{a priori} when fitting with \SPARK. The over-absorption (and thus overestimated opacities) at 3.4 and 4.4 days comes especially from barium (${}_{56}$Ba) at $\lesssim$$5500$~\AA~and cerium at $\sim$$6000-7500$~\AA. We in turn find overly large Planck mean opacities, and later poor light curves and unphysical ejecta masses, unless we exclude altogether the expansion opacity at $\leqslant4500$~\AA~at 1.4 and 2.4 days and $\leqslant6400$~\AA~at 3.4 and 4.4 days. These cutoffs are based on visual inspection of where our spectral fit begins to perform poorly. It is not surprising that a wavelength-cutoff is needed given the systematic lack of atomic lines for most $r$-process elements, and especially the lanthanides and actinides, which have high opacities in this spectral range. Indeed, most spectral modeling that relies on radiative transfer (\eg, \citealt{gillanders22}) also struggles at these shorter wavelengths. The opacities with wavelength-cutoffs are thus used for the remainder of this work. We discuss the choice of cutoffs further in Section~\ref{ssc:disco-opacities}.

To verify that imposing a cutoff at the step of computing Planck mean opacities does not introduce any inconsistencies at 1.4 and 2.4 days, we rerun \SPARK~at these epochs with these $\leqslant4500$~\AA~cutoffs in the inference, masking these wavelengths when computing likelihoods. In our spectral fits at 3.4 and 4.4 days, we already exclude spectral ranges at wavelengths of $\leqslant6400$~\textit{a priori}. These new inferences at 1.4 and 2.4 days yield best-fit parameters, abundances, heating rates, and opacities which are consistent with our original fits, and the best-fit light curves later in our analysis show no substantial differences using these alternate opacities.

Figure~\ref{fig:hists-opacities} shows the inferred posterior probability distributions for these opacities computed using a wavelength-cutoff. We tabulate the ingredients (velocity, density, temperature, and wavelength-cutoff) that go into the opacity calculations and the resulting opacities in Appendix~\ref{app:opacities}, and compare these results to previous assumptions used in the literature in Section~\ref{ssc:disco-opacities}. We again caution that these distributions are not distributions describing the physical conditions in the ejecta, but rather probabilistic distributions describing the best-fit parameters. We also emphasize that the total opacities at 3.4 and 4.4 days are not simply mass-weighted sums of the two components' opacities. Instead, they are obtained by mass-weighting and summing both components' abundance patterns \textit{a priori}, and then computing the opacity of this mixture of two components. The single-component fits yield simple, unimodal posteriors at 1.4 and 2.4 days, while the multi-component fits yield more complex posteriors at 3.4 and 4.4 days. The posterior distribution for the total opacity at 3.4 days presents a broad range of solutions, extending from $\sim$$3 - 30~\mathrm{cm^2~g^{-1}}$. At 4.4 days, the posterior for the opacity is peaked at $3.9~\mathrm{cm^2~g^{-1}}$ but then displays a tail of possible solutions extending to high opacities of $\sim$$20~\mathrm{cm^2~g^{-1}}$. These opacities can be better understood by examining the opacities of the individual components. Figure~\ref{fig:hists-opacities}~also shows the posterior distributions of the Planck mean opacities of each component, computed independently. At 3.4 days, the blue component has $Y_{e} = 0.228^{+0.073}_{-0.088}$, nearly centered on the oft-quoted $Y_e \sim 0.25$~threshold between red and blue ejecta, yielding a broad opacity posterior. The red component has $Y_{e} = 0.161^{+0.149}_{-0.104}$, more firmly in the red, but displaying a tail in the blue, which is reflected by the posterior distribution for the opacity of this component being peaked at $16.8~\mathrm{cm^2~g^{-1}}$ but extending down to $\sim$$2~\mathrm{cm^2~g^{-1}}$. Combining these two components into a mixture which is strictly neither red nor blue thus yields a broad range in possible opacities in the posterior. At 4.4 days, the blue and red components have $Y_{e} = 0.280^{+0.098}_{-0.162}~\mathrm{and}~0.209^{+0.116}_{-0.113}$, respectively. The blue component then shows a strong peak in possible solutions at low opacities of $4.2~\mathrm{cm^2~g^{-1}}$, with a moderate tail to higher opacity solutions of $\sim$$30~\mathrm{cm^2~g^{-1}}$, while the red component shows two distinct modes in the posterior at $\sim$$4~\mathrm{cm^2~g^{-1}}$ and $\sim$$30~\mathrm{cm^2~g^{-1}}$. Combining these two components into a mixture with a marginally lower abundance of lanthanides compared to that at 3.4 days (see Figure~\ref{fig:bestfit-abundances}) yields a posterior distribution with a peak at lower opacities but a range of possibilities at higher opacities.

Finally, Figure~\ref{fig:planck-opacities} shows these total opacities at 1.4, 2.4, 3.4, and 4.4 days, as a function of velocity. Recall that we only infer the presence of the red component as of 3.4 days, but the red component extends out to $v_{\mathrm{outer}} \sim 0.35c$ (the same as the blue component at 1.4, 2.4, 3.4, and 4.4 days) when detected at 3.4 and 4.4 days. Indeed, at 3.4 and 4.4 days, we see no evidence for a radial gradient in the composition from $v = 0.17c$~to~$0.35c$~based on our inference with \SPARK. In \vieiratwofour, we interpreted our detection of the red ejecta component at 3.4 days as the emergence of a red component that was always present, but was previously outshined by a blue component at earlier epochs. Thus, aside from the difference in composition at 1.4 and 2.4 days versus 3.4 and 4.4 days, the opacity as a function of velocity is also largely dictated by the density and temperature profile. The ejecta density in \TARDIS~scales as $\rho \propto v^{-3}$, while expansion opacity follows $\kappa_{\mathrm{exp}} \propto \rho^{-1}$, such that opacities decrease at smaller velocities (higher densities) at 2.4, 3.4, and 4.4 days. This effect competes with the temperature dependence: the Planck mean opacity uses the Planck function at some $T$ to weight the opacities\footnote{A blackbody peaks at wavelengths $\sim$5330, 7360, 8480, and 8700~\AA~for temperatures of $5440,~3940,~3420,~\mathrm{and}~3330~\mathrm{K}$, respectively.} and the higher temperature ($5440~\mathrm{K}$) at 1.4 days changes the ionization fraction of the ejecta. The ejecta is also most lanthanide-poor at 1.4 days. This combination of high temperatures and a dearth of high-opacity lanthanides yields the lowest opacity among all epochs at 1.4 days. We compare these inferred gray opacity values to those previously adopted in the literature in Section~\ref{ssc:disco-opacities}. 

We emphasize again that it is non-trivial that the opacities inferred in the line-forming region should be applicable to the ejecta contained at smaller velocities. Indeed, because our inferred $v_{\mathrm{inner}}$~with $W_{\mathrm{inner}} \sim 0.3 < 0.5$~lies above the photosphere, our inferred opacity does not represent the entire photospheric region. This point is especially relevant because the opacity has a density dependence. We use these opacities in a light curve model to explicitly assess the ability to extend our quantities inferred in the line-forming region to the material outside the line-forming region and even below the photosphere, and, to explicitly test our ability to connect spectral and time-domain modeling.

\subsection{Light Curve Model and Inputs}\label{ssc:lightcurve-model}

We use the light curve model of \cite{hotokezaka20}.\footnote{We specifically adapt \texttt{kilonova-heating-rate:\newline\href{https://github.com/Basdorsman/kilonova-heating-rate}{https://github.com/Basdorsman/kilonova-heating-rate}}} This 1-dimensional model computes the luminosity output by homologously-expanding spherically-symmetric ejecta that is heated by radioactive decays, producing thermal emission. The ejecta is characterized by some number of concentric radial shells, each at its own density $\rho$, velocity $v$, and opacity $\kappa$. Some total mass $M_{\mathrm{ej}}$ is distributed according to a time- and velocity-dependent power-law density profile, $\rho(t, v) = \rho_0 (t / t_0)^{-3} (v / v_0)^{\beta}$, where $v = r/t$ at all radii $r$ and times $t$ under homologous expansion. Energy is injected into the shells according to the radioactive heating rate $\dot{q}(t)$, multiplied by a thermalization efficiency $\epsilon_{\mathrm{th}}(t)$~which describes how efficiently the energy contained in radioactive decay products is deposited into the ejecta. The luminosity of all shells is computed and summed to produce the output light curve. This setup of several concentric shells is analogous to the shells used in \TARDIS. We opt for this model, with its significant simplification in the form of gray opacities, as it is typical of those used in photometric modeling of kilonovae. Below, we describe which of the model parameters are set according to our inference on the spectra, which parameters are fixed, and which parameters are fit. 

We set up an array of velocities from a minimum $v_{\mathrm{min}} = v_0 = 0.1c$ to a maximum $v_{\mathrm{max}} = 0.4c$ in all models. These extend beyond the $v_{\mathrm{inner}}$ and $v_{\mathrm{outer}}$ that we infer from spectral modeling. This choice is intentional: while spectral modeling is only sensitive to the ejecta in the line-forming region, light curve modeling probes beyond the boundaries of this line-forming region and below the photosphere. Our particular spectral modeling, where the inferred $v_{\mathrm{inner}}$~lies above the photosphere, does not capture the entire photospheric region of the ejecta. Nonetheless, this setup of ejecta spanning several stratified shells is similar to that used for spectral synthesis in \TARDIS. We fix $\beta = -4.5$ in our power-law density profile, as is done in \cite{hotokezaka20}. The choice of density profile affects the radial profile of the thermalization efficiency and thus the slope of the light curve, but weakly so. We re-perform our later light curve fits with $\beta = -6.0, -5.5, -5.0, ..., -2.5$~and find variations in the inferred ejecta mass of $\lesssim 10\%$ alongside variations in the bolometric (multi-band) light curves of $\pm \sim$$\mathrm{10^{41}~\mathrm{erg~s^{-1}}}$ ($\pm \sim$$0.2~\mathrm{AB~mag}$), well within the uncertainties introduced by the uncertainties in opacities and heating rates.

For the heating rates, we use our $t^{-0.99}$ power-law fit to our inferred heating rates (Section~\ref{ssc:infer-heatingrates}, Figure~\ref{fig:points-heatingrates}) to describe $\dot{q}(t)$~as a smooth function of time. To describe how much of this heating actually powers the kilonova, we also require a thermalization efficiency $\epsilon_{\mathrm{th}}$. We use the mass and radially dependent (equivalently, velocity-dependent) thermalization efficiency $\epsilon_{\mathrm{th}}(t, v)$ described in \cite{wollaeger18}, which takes into account thermalization of the products of alpha decays, beta decays, gamma decays, and fission. Following \cite{wollaeger18}, we partition the energy carried by the products of different decays as: $\alpha$~particles at 5\%, $\beta$~particles (electrons/positrons) at 20\%, gamma-rays at 40\%, and fission fragments at 0\%, while the remaining 35\% is carried away by neutrinos. The mass and radial dependence of the thermalization efficiency come from the dependence of $\epsilon_{\mathrm{th}}$~on $\rho(t, v)$, and so each shell in the velocity array is then described by its own unique thermalization efficiency. Given the wide range in masses we explore in this work and our array of shells at different velocities, this general approach is more appropriate than the oft-used \cite{barnes16} thermalization efficiency, which is provided in analytic form for only masses in the range $0.001$~to $0.05~M_{\odot}$~and is only dependent on time. Moreover, \cite{brethauer24} highlight the important biases in ejecta mass estimates which are introduced by assuming a global thermalization efficiency without radial dependence.

We also provide the gray opacity in each shell, as required by the light curve model. In practice, to fit the suite of light curve data from the kilonova, we require a light curve model that is smooth in time and can be evaluated outside of the discrete times at which we perform spectral inference. We therefore also require a smooth opacity $\kappa(t, v)$. Because our spectral modeling is sensitive only to the material in the range $(v_{\mathrm{inner}},~v_{\mathrm{outer}})$, we only infer the opacities in these velocity ranges at each epoch (Section \ref{ssc:infer-opacities}, Figure~\ref{fig:planck-opacities}). As the ejecta expands, the photosphere recedes, decreasing the inferred $v_{\mathrm{inner}}$ and enabling measurement of the opacity deeper into the ejecta, but there remains some fraction of the ejecta which is unseen. Indeed, because our inferred $v_{\mathrm{inner}}$~does not correspond to the photosphere, we cannot claim to measure the opacity (or any of the quantities we infer) in the entirety of the photospheric region, let alone the entirety of the ejecta. Fitting in velocity is complicated by our not knowing these opacities at smaller velocities, where most of the ejecta mass lies, especially at 1.4 and 2.4 days. Furthermore, although we infer multiple components in the ejecta, our spectral fits show no evidence for strong radial dependence in composition from $0.17c~\mathrm{to}~0.35c$~for these components. Finally, we only have two data points for the opacity of a red component. 

Given these constraints, we opt for a purely time-dependent fit to the total opacity. We include the mixture of red and blue components at 3.4 and 4.4 days, which is flat in velocity at all times. We fit the opacities with a power-law $\kappa(t) = \kappa_0 (t / ~\mathrm{1~day})^{\gamma}$, finding $\kappa_0 = 1.17 \pm 0.18~\mathrm{cm^2~g^{-1}}$ and $\gamma = 1.4 \pm 0.4$. We show the time-dependent opacities and our power-law fit in Figure~\ref{fig:planck-opacities-fitt}. Time-dependent opacities have been used elsewhere, \eg, \cite{waxman18} find a power-law opacity $\kappa \propto t^{\gamma}$ with $0.5 \lesssim \gamma \lesssim 1$, peaked at $\gamma \approx 0.6$ by computing bolometric luminosities from the multi-band data and fitting these luminosities. 

Given our fixed velocities $v_0 = v_{\mathrm{min}}$ and $v_{\mathrm{max}}$, our fixed $\beta$ in the density power-law, our \SPARK-inferred and fit heating rates $\dot{q}(t)$, our chosen \cite{wollaeger18} thermalization efficiency $\epsilon_{\mathrm{th}}(t)$, and our \SPARK-inferred and fitted opacity $\kappa(t)$, the only remaining model parameter is then $M_{\mathrm{ej}}$, the total mass of the ejecta between $v_0$ and $v_{\mathrm{max}}$. We caution that this is not the mass we report in \vieiratwothree~or \vieiratwofour; the spectral modeling in those works is only sensitive to the mass in the line-forming region, and our inferred $v_{\mathrm{inner}}$~lies above the photosphere. Indeed, given the steep power-law density profile of the ejecta, the vast majority of the ejecta mass should be located below $v_{\mathrm{inner}}$, particularly at early times. We thus fit the light curve data of the GW170817 kilonova to infer this total ejecta mass.

\subsection{Light Curve Inference on the GW170817 Kilonova}\label{ssc:lightcurve-fitting}

We use \texttt{redback} (\citealt{sarin24a}) to collect and fit the photometric data of the GW170817 kilonova and infer the total ejecta mass, $M_{\mathrm{ej}}$. We fit the bolometric luminosities as computed in \cite{waxman18}, in which they convert apparent magnitudes to fluxes and perform trapezoidal integration of those fluxes with an added Rayleigh-Jeans ($\lambda^{-4}$) tail at the longest wavelengths. Within \texttt{redback}, we use \texttt{bilby} (\citealt{ashton19}) to perform nested sampling with \texttt{dynesty} (\citealt{speagle20}). We impose a log-uniform prior on $M_{\mathrm{ej}}$ from $0.005~M_{\odot}$ to $0.500~M_{\odot}$.

We also use \texttt{redback} to obtain data in the $u, g, r, i, z, J, H,~\mathrm{and}~K$ bands\footnote{For this work, we denote as  `$K$' both the $K$-band and $K_s$ band, since $K_s$  observations at $<$$1$~day provide key constraints on the bolometric corrections.}, from the Open Access Catalogue (\citealt{guillochon17}), for comparison of our best fit to the multi-band data. To convert the bolometric luminosity of our model into band magnitudes, we compare the bolometric luminosities of \cite{waxman18} to the multi-band data of the kilonova, to obtain bolometric corrections. To obtain a smooth set of bolometric corrections as a function of time, we fit a Gaussian Process (GP) to the bolometric corrections, independently in each band. 

We emphasize that the shape and slope of our model light curves are dictated primarily by our input opacities and heating rates, while the overall light curve amplitude is set by the total ejecta mass. Thus, the shape of our model light curves is informed by our spectral modeling, and the light curve modeling here fits the light curve amplitude.


\begin{figure}[ht!]
    \centering
    \includegraphics[width=0.95\linewidth]{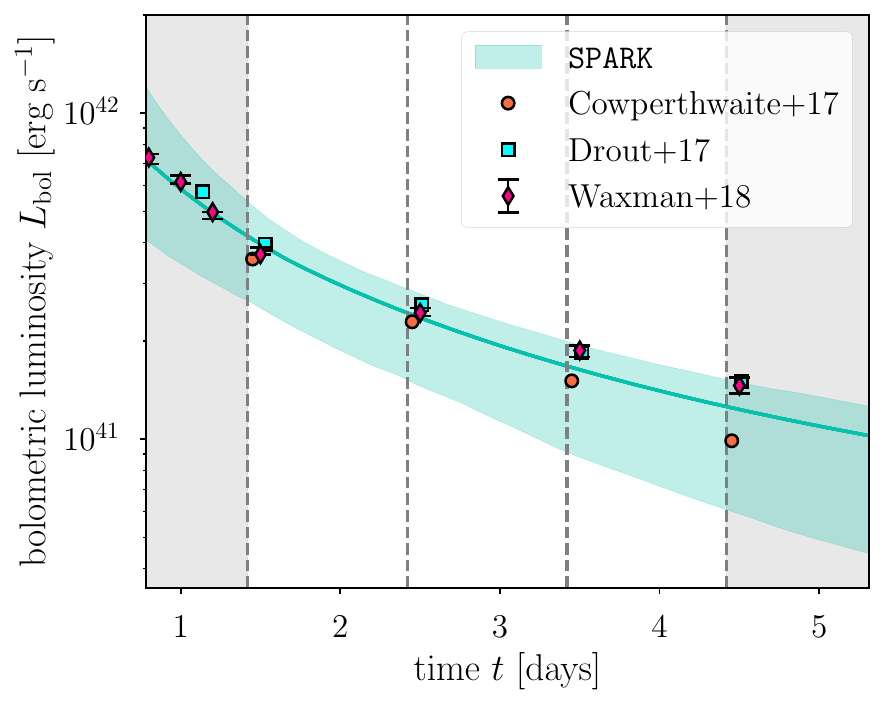}
    \caption{\textbf{Best-fit bolometric light curves for the GW170817 kilonova, with input heating rates and opacities informed by spectral inference with \SPARK}. We fit the bolometric luminosities of \cite{waxman18}, but also show the bolometric luminosities from two other calculations (\citealt{cowperthwaite17, drout17}) for comparison. The shaded green band spans the 2.5\% and 97.5\% percentiles of 1000 samples from our \SPARK~posterior for the heating rates and opacities, for fixed best-fit mass, reflecting the impact of those parameters' uncertainties on the light curve. Gray dashed vertical lines are epochs at which we perform inference with \SPARK. Gray shaded regions at $< 1.4~\mathrm{days}$~($>4.4~\mathrm{days}$) demarcate times before (after) our earliest (latest) spectral inference: at these times, our model light curves rely on extrapolation of our inferred heating rates and opacities, and are not fit to the observed datapoints.}
    \label{fig:lightcurves-bolometric}
\end{figure}

\begin{figure*}[ht!]
    \centering
    \includegraphics[width=0.9\linewidth]{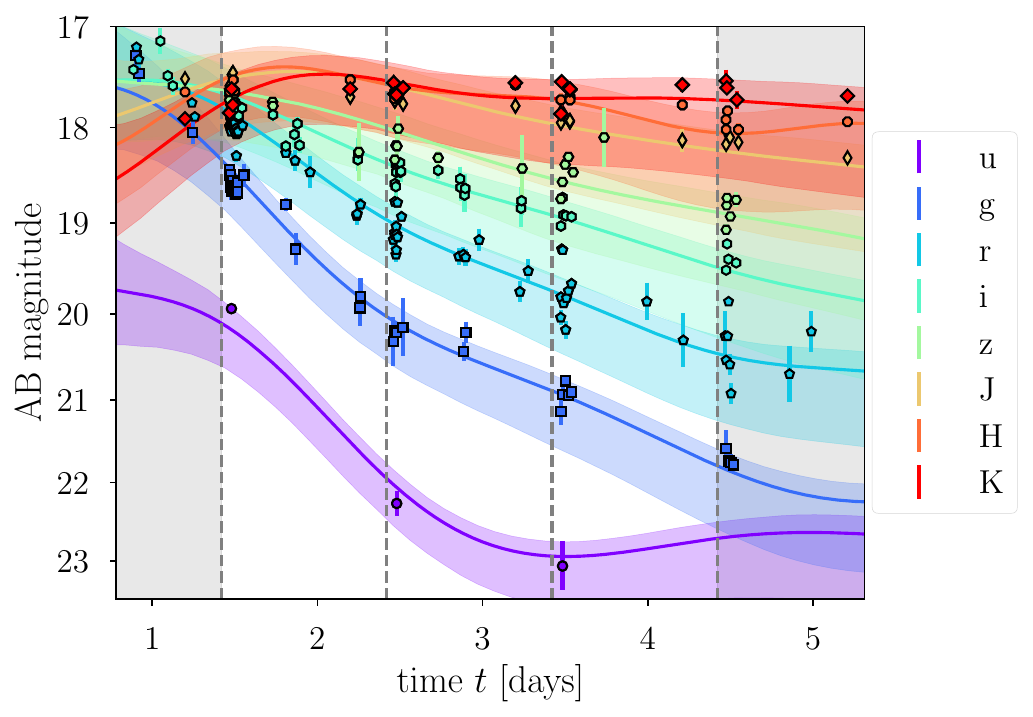}
    \caption{\textbf{Multi-band light curves for the GW170817 kilonova, corresponding to the best-fit bolometric luminosity in Figure~\ref{fig:lightcurves-bolometric}, with input heating rates and opacities informed by spectral inference with \SPARK.} Bolometric corrections are obtained from \cite{waxman18}. We include the $u$, $g$, $r$, $i$, $z$, $J$, $H$, and $K$-band data, collected from the Open Access Catalogue (\citealt{guillochon17}) with \texttt{redback}. Markers indicate the data, and solid lines show the best fit. Shaded bands span the 2.5\% and 97.5\% percentiles of 1000 samples from our \SPARK~posterior for the heating rates and opacities, for fixed best-fit mass, reflecting the impact of those parameters' uncertainties on the light curves. Gray dashed vertical lines are epochs at which we perform inference with \SPARK. Gray shaded regions at $< 1.4~\mathrm{days}$~($>4.4~\mathrm{days}$) demarcate times before (after) our earliest (latest) spectral inference: at these times, our model light curves rely on extrapolation of our inferred heating rates and opacities, and are not fit to the observed datapoints.}
    \label{fig:lightcurves}
\end{figure*}

\begin{figure*}[ht!]
    \centering
    \includegraphics[width=0.9\linewidth]{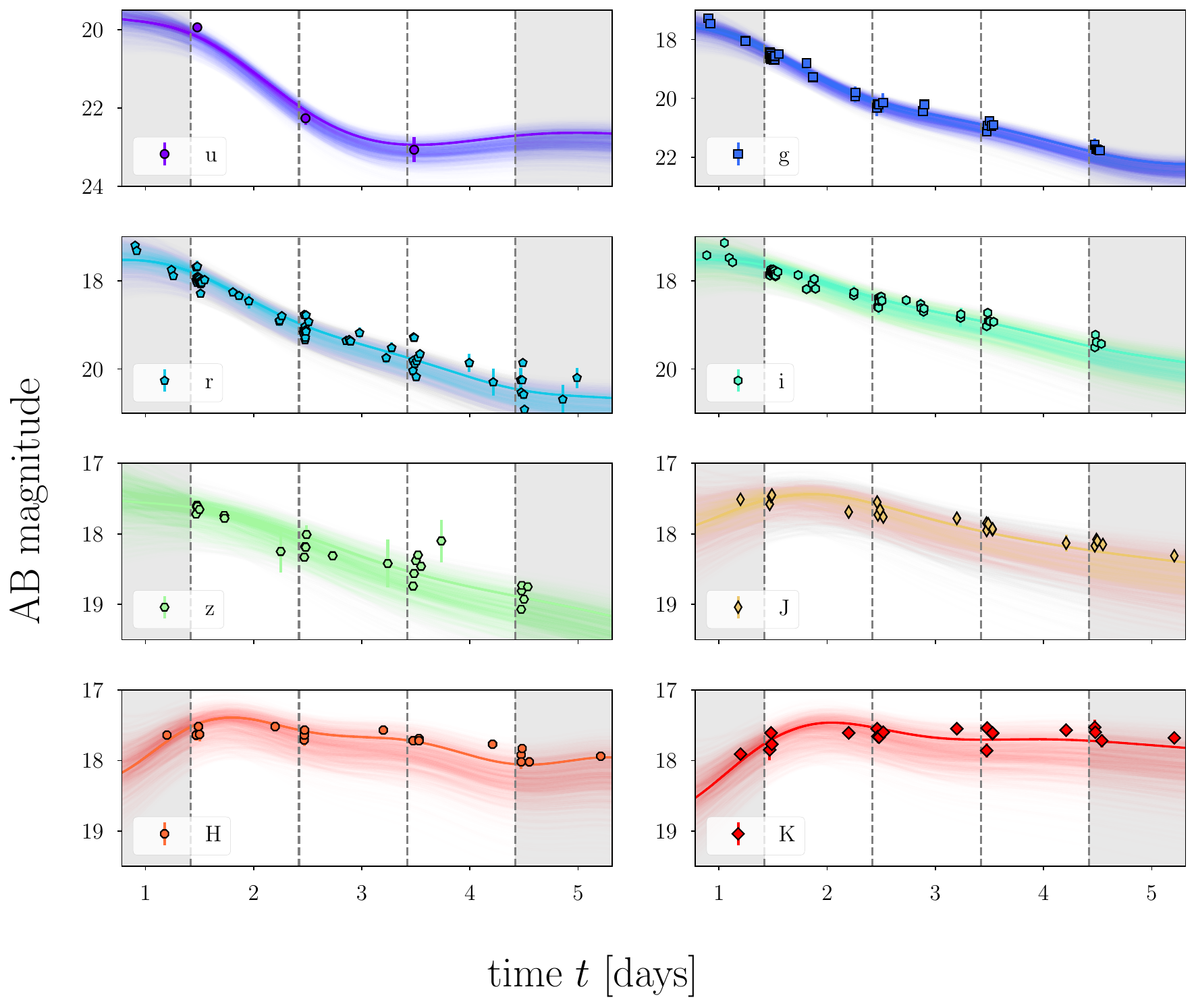}
    \caption{\textbf{Multi-band light curves for the GW170817 kilonova, corresponding to the best-fit bolometric luminosity in Figure~\ref{fig:lightcurves-bolometric}, with input heating rates and opacities informed by spectral inference with \SPARK.} Bolometric corrections are obtained from \cite{waxman18}. We include the $u$, $g$, $r$, $i$, $z$, $J$, $H$, and $K$-band data, collected from the Open Access Catalogue (\citealt{guillochon17}) with \texttt{redback}. Markers indicate the data and solid lines show the best fit. Translucent lines are 1000 samples from our \SPARK~posterior for the heating rates and opacities, for fixed best-fit mass, reflecting the impact of those parameters' uncertainties on the light curves. Gray dashed vertical lines indicate epochs at which we perform spectral inference. Gray shaded regions at $< 1.4~\mathrm{days}$~($>4.4~\mathrm{days}$) demarcate times before (after) our earliest (latest) spectral inference: at these times, our model light curves rely on extrapolation of our inferred heating rates and opacities, and are not fit to the observed datapoints. Limits on the magnitude axis vary for different bands for legibility.}
    \label{fig:lightcurves-multiband}
\end{figure*}

\section{Results}\label{sec:results}

\subsection{Best-Fit Light Curves}\label{ssc:results-lightcurves}

Figure~\ref{fig:lightcurves-bolometric}~shows our best fit light curves to the bolometric luminosities of the GW170817 kilonova. Figures~\ref{fig:lightcurves}~and~\ref{fig:lightcurves-multiband} show the corresponding multi-band light curves after applying the bolometric corrections described above. We also plot (but do not fit) data just prior to 1.4 days ($\sim 0.8 - 1.4~\mathrm{days}$) and after 4.4 days ($\sim 4.4 - 5.4~\mathrm{days}$), to visually examine the ability of our model to match the observed light curve immediately outside of the time range where we have spectral inference with \SPARK. The best-fitting light curve is described by a total ejecta mass of $M_{\mathrm{ej}} = 0.110^{+0.003}_{-0.003}~M_{\odot}$~between $0.1c$~and~$0.4c$. This uncertainty on the inferred mass is underestimated; in Section~\ref{ssc:results-masses}, we discuss and provide a more reasonable estimate of the uncertainties by incorporating the uncertainties on heating rates and opacities, and find $M_{\mathrm{ej}} = 0.11^{+0.04}_{-0.02}~M_{\odot}$. The inferred total ejecta mass and the quality of the light curve fit are not sensitive to the choice of $v_{\mathrm{min}}$, with the fit yielding similar masses $\pm$ $\sim$$0.001~M_{\odot}$~and similar light curves for variations in the range of $0.05c - 0.15c$. 

The total ejecta mass sets only the amplitude of the light curve, while the input heating rates and opacities dictate its shape. To capture the impact of heating rates and opacities on the shape of the light curves, we recompute our best fit light curve for fixed ejecta mass but variable heating rates and opacities, sampled from our \SPARK~posteriors. This sampling of heating rates and opacities introduces an uncertainty of approximately $\pm \sim$$1 -  4\times10^{41}~\mathrm{erg~s^{-1}}$ ($\pm \sim$$0.5 - 1~\mathrm{mag}$)~in the bolometric (multi-band) light curves (see shaded bands in Figures~\ref{fig:lightcurves-bolometric}~and~\ref{fig:lightcurves}~and~traces in Figure~\ref{fig:lightcurves-multiband}). This sampling shows the impact of these quantities on the shapes of the light curves.

Overall, given the smooth input heating rates, opacities, and thermalization efficiency, the best-fit bolometric light curve is smooth in time in all bands. The multi-band light curves also exhibit expected chromatic behavior, peaking at earlier times in the bluer bands, and then later in the redder bands, as a direct consequence of our applied bolometric corrections (see Section ~\ref{ssc:lightcurve-fitting}). The redder emission then persists, while the blue emission quickly fades. One exception is our $u$-band light curve, where the light curve unexpectedly plateaus after $\sim$3.4 days. This is because the $u$-band light curve is the most sparse, with only 3 data points from 1.4 to 3.4 days. As a result, the bolometric corrections are not constrained beyond 3.4 days, and our Gaussian Process for these corrections predicts an unphysical flat bolometric correction to the $u$-band beyond 3.4 days. 

In addition to fitting the bolometric luminosity of \cite{waxman18}, our light curve is broadly consistent with other calculations for the bolometric luminosity (\eg, \citealt{cowperthwaite17, drout17}). We include these other calculations in Figure~\ref{fig:lightcurves-bolometric}. If we instead fit the bolometric light curves estimated in those studies, we infer ejecta masses within $\pm10\%$ of the fit to the \cite{waxman18} luminosities, and obtain nearly identical light curves. Our model light curves also match the observed data immediately before 1.4 days ($\sim 0.8 - 1.4~\mathrm{days}$) and beyond 4.4 days ($\sim 4.4 - 5.4~\mathrm{days}$), despite excluding this data from our light curve inference because our earliest spectral inference is at 1.4 days and our latest at 4.4 days.

\subsection{Inferred Total Ejecta Mass}\label{ssc:results-masses}

\begin{figure}[ht!]
    \centering
    \includegraphics[width=0.95\linewidth]{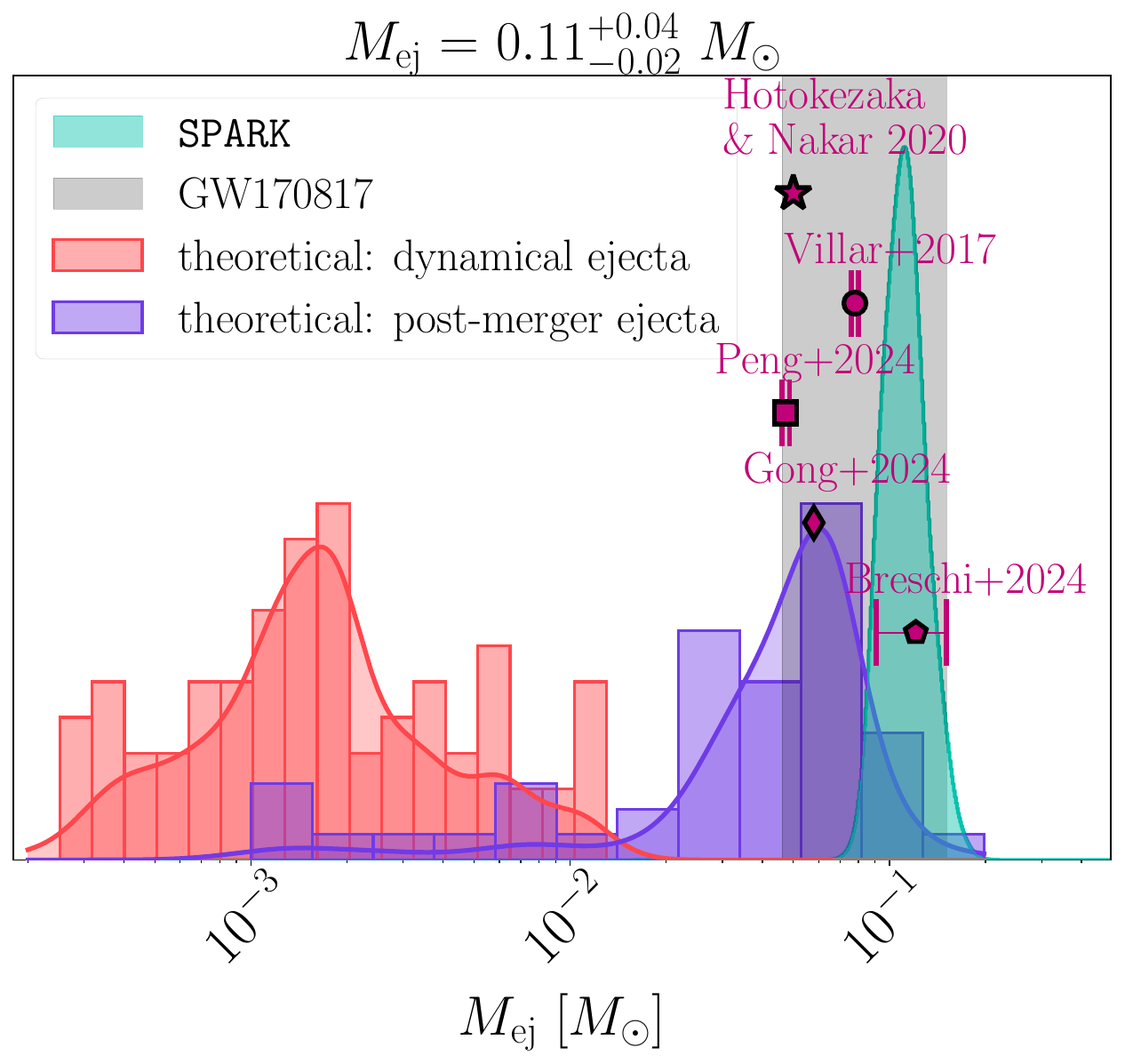}
    \caption{\textbf{Total ejecta mass of the the GW170817 kilonova, inferred using a light curve model informed by heating rates and opacities inferred with \SPARK.} We infer a mass of $M_{\mathrm{ej}} = 0.11^{+0.04}_{-0.02}~M_{\odot}$. For comparison, we include the masses measured with various analytic and radiative transfer fits to GW170817 (magenta points). Uncertainties, included as error bars when available, are small relative to the full extent of parameter space. The shaded gray region shows the span of these various measurements. We also show a sample of dynamical ejecta masses (red) and post-merger ejecta masses (purple-blue) from theoretical GRMHD / numerical relativity simulations for a variety of NS-NS / NS-BH systems. The heights of these theoretical mass distributions and the vertical positions of the measured masses carry no meaning and are adjusted for legibility. See Section \ref{ssc:results-masses} for details on the origins of these models' and simulations' masses.}
    \label{fig:mass}
\end{figure}

Figure~\ref{fig:mass} shows our posterior for the total ejecta mass, in comparison to measurements from other works using a variety of different approaches, as well as an ensemble of theoretical results. Without propagating uncertainties in the heating rates and opacities into the total ejecta mass, we find $M_{\mathrm{ej}} = 0.11$~between $0.1c$~and~$0.4c$, but severely underestimate the uncertainties. A more robust method to propagating uncertainties could be a Monte Carlo approach, in which we sample heating rates and opacities from our \SPARK~posteriors and perform new light curve fits, repeating for many samples to obtain a distribution of masses. However, this is computationally infeasible due to the time required for a single light curve fit. To approximately estimate the uncertainty on the total ejecta mass, we re-perform the light curve fit for the minimum (2.5\% percentile) and maximum (97.5\% percentile) of the heating rates and opacities. Varying only the opacities within these percentiles yields a range of masses of $\sim$$0.098 - 0.143~M_{\odot}$, while varying only the heating rates within their percentiles yields a range of masses of $\sim$$0.094 - 0.136~M_{\odot}$. Finally, varying both opacities and heating rates within their quantiles yields a range of masses of $\sim$$0.082 - 0.177~M_{\odot}$. As expected, larger (smaller) input opacities and/or smaller (larger) heating rates yield larger (smaller) inferred ejecta masses. Large masses of $\sim$$0.177~M_{\odot}$~are only possible using both the lowest heating rates and the highest opacities allowed by our \SPARK~posterior probability distributions, while small masses of $\sim$$0.082~M_{\odot}$ are only possible using both the lowest opacities and highest heating rates. Given these ranges, we adopt the more conservative estimate of the total ejecta mass and its associated uncertainty of $0.11^{+0.04}_{-0.02}~M_{\odot}$.

We compare our inferred total ejecta mass to results from a diverse variety of other approaches, including models with various dimensionality (1, 2, or 3D kilonovae), both analytic and semi-analytic models, and results from radiative transfer simulations. There are indeed a wealth of measurements of the ejecta masses of GW170817 (see \citealt{siegel19_GW170817, ji19} and references therein); here, we choose four measurements in particular. \cite{hotokezaka20}---from which we draw our analytic light curve model---are able to reasonably describe the light curves with $0.05~M_{\odot}$ of ejecta, with low opacities ($0.5~\mathrm{cm^2~g^{-1}}$) at high velocities ($0.2c < v \leqslant 0.4c$) and moderate opacities ($3~\mathrm{cm^2~g^{-1}}$) at lower velocities ($0.1c \leqslant v \leqslant 0.2c$), with a power-law density $\propto v^{-4.5}$. \citet{villar17} fit the multi-band light curves with an analytic, 1D, three-component blue (opacity $0.5~\mathrm{cm^2~g^{-1}}$) + purple ($3~\mathrm{cm^2~g^{-1}}$) + red ($10~\mathrm{cm^2~g^{-1}}$) ejecta model and find a total mass $M_{\mathrm{blue}} + M_{\mathrm{purple}} + M_{\mathrm{red}} = 0.020^{+0.001}_{-0.001} + 0.047^{+0.001}_{-0.002} + 0.011^{+0.002}_{-0.001} = 0.078^{+0.002}_{-0.002}~M_{\odot}$. \citet{peng24} perform fitting via a neural network light curve emulator based on 2D radiative transfer simulations with the \texttt{SuperNu}~code, and find a total dynamical ejecta (redder, $Y_e = 0.04$) + wind ejecta (bluer, $Y_e = 0.27$) with a mass of $M_{\mathrm{d}} + M_{\mathrm{w}} = 0.0192^{+0.0010}_{-0.0010} + 0.0260^{+0.0006}_{-0.0006} = 0.0452^{+0.0013}_{-0.0013}~M_{\odot}$. \citet{gong24} extend the analytic 3D, three-component model of \citet{zhu20} to include an analytic fit for the wavelength-dependent opacity of each of three components: the dynamical ejecta, a neutrino-driven wind, and a viscously-driven wind. They find $M_{\mathrm{d}} + M_{\mathrm{n}} + M_{\mathrm{v}} = 0.018 + 0.020 + 0.020 = 0.058~M_{\odot}$. They are unable to quantify uncertainties with their technique, but find that these parameters reproduce the observations. Finally, \citet{breschi24} perform a multi-messenger fit which incorporates both kilonova parameters and parameters of the BNS from gravitational wave data, including the chirp mass, mass ratio, and spin of the binary. They infer a dynamic + wind $M_{\mathrm{d}} + M_{\mathrm{w}} = 0.001^{+0.001}_{-0.003} + 0.12^{+0.03}_{-0.02} = 0.121^{+0.030}_{-0.020}~M_{\odot}$, larger than that inferred from electromagnetic observations alone. 

Our inferred total ejecta mass is towards the higher-mass end of the range spanned by these previous studies. Notably, the mass we infer is larger than that found in \citet{hotokezaka20}, from which our model derives. This higher mass arises primarily from the larger opacities we infer from the spectra relative to those assumed in \citet{hotokezaka20}, and the known degeneracy between ejecta mass and opacity. Averaged over time, we infer an opacity $\sim$$6~\mathrm{cm^2~g^{-1}}$, twice as large as the opacity of $3~\mathrm{cm^2~g^{-1}}$~that \cite{hotokezaka20}~assume for $0.1 \leqslant v \leqslant 0.2c$. Our inferred mass is then twice as large as that of \cite{hotokezaka20}, as expected. Our larger inferred opacity may be a consequence of the fact that our inner boundary velocity $v_{\mathrm{inner}}$~in our spectral modeling lies above the photosphere, such that we infer the opacity in only a small fraction of the ejecta. Interestingly, our inferred total ejecta mass is in agreement with that found by \citet{breschi24} (which albeit has the largest uncertainties). That work was informed not only by light curves, but also by parameters inferred from gravitational-wave data. 

In Figure~\ref{fig:mass}, we also compare to the masses of dynamical ejecta and post-merger ejecta borne out of general relativistic magneto-hydrodynamic (GRMHD) and numerical relativity (NR) calculations. These calculations (\citealt{radice18,fujibayashi20,just22accr,fahlman18,combi23,fahlman23,fujibayashi23,just23,christie19,fernandez19,fahlman22,just22dyn,zappa23}) encompass merger/post-merger simulations with different progenitor masses, mass ratios, fates of the merger remnant, treatments of magnetic fields, treatments of viscosity, and treatments of weak interactions through neutrinos. We collect a total of 82 dynamical mass calculations and 48 post-merger mass calculations. These dynamical ejecta encompass the tidal ejecta expelled due to tidal disruption of the neutron stars and/or the shock-heated ejecta expelled along the poles of the merger due to the contact between two neutron stars. The post-merger ejecta includes outflows from an accretion disk around a remnant black hole/neutron star and/or winds from the surface of a remnant hypermassive neutron star. This post-merger ejecta is more massive, but displays a lower-mass tail which overlaps with dynamical ejecta. 

Our large inferred total ejecta mass of $0.11^{+0.04}_{-0.02}~M_{\odot}$~is difficult to explain without substantial contributions from post-merger ejecta. Such post-merger ejecta can be blue, in the case of winds from a hypermassive neutron star remnant (\citealt{curtis24, kiuchi24}) or a heavily neutrino-irradiated accretion disk wind (\citealt{hayashi22, just22accr}), or red, in the case of a neutron-rich, magnetically-driven accretion disk wind (\citealt{christie19}). We discuss the implications of this large ejecta mass, combined with the insights gained from our spectral inference, in the following sections. 


\section{Discussion}\label{sec:disco}

\subsection{Opacities of the Ejecta}\label{ssc:disco-opacities}

Using the opacities inferred from our spectral modeling with \SPARK~to perform light curve modeling fundamentally links the spectral and time domains. However, a fundamental limitation of our spectral analysis is that our inferred inner boundary velocity $v_{\mathrm{inner}}$~lies above the photosphere, and our spectral modeling only captures a small fraction of the total ejecta. Thus, we must first assess whether our inferred opacities are realistic and consistent with results from systematic opacity calculations, as well as typical values assumed in light curve modeling alone.

Our best-fit inferred opacities, ranging from smallest $1.9^{+0.2}_{-0.2}~\mathrm{cm^2~g^{-1}}$ at 1.4 days to largest $9.6^{+18.5}_{-6.4}~\mathrm{cm^2~g^{-1}}$ at 3.4 days, match the effective gray opacities from systematic opacity calculations. In particular, our opacities closely resemble the Planck mean opacities of \cite{tanaka20}, which are obtained using systematic opacity calculations for elements from $Z = 26$~to~$88$~for a wide range of relevant temperatures. Moreover, \cite{tanaka20} employ the same reaction network calculations (\citealt{wanajo18}) currently used in \SPARK~to map $Y_e$~to the composition of the ejecta. Our inferred gray opacities best match their theoretical calculations for~$Y_e \sim 0.15,~0.20,~0.25,~0.30,~0.35$~(see their Figure 11), as expected given the $Y_e$~we infer for red and blue components from spectral modeling. This agreement between the gray opacities is noteworthy given some disagreement between the wavelength-dependent expansion opacities. Beyond our imposed wavelength cutoffs, our expansion opacities 
(Figure~\ref{fig:expansion-opacities}) drop more sharply with wavelength than those of theoretical calculations, hinting at the incompleteness of our line list. Yet, our Planck mean opacities closely match those of \cite{tanaka20}. We ascribe the agreement between our Planck mean opacities and those of more complete theoretical line lists to the fact that Planck mean opacity is computed by weighting the expansion opacity by the Planck function $B_{\lambda}(T)$, which peaks at wavelengths $\lambda \sim5300-8700$~\AA~for the temperatures of the ejecta $T \sim 3300 - 5400~\mathrm{K}$ from 1.4 to 4.4 days. At these wavelengths of greatest importance in computing Planck means, the deficits in our inferred expansion opacities are less severe. Thus, despite any line list incompleteness, we obtain the reasonable gray opacities input to our light curve model. The deficit in our expansion opacities at the longest wavelengths should become more impactful on the gray opacity at later epochs, when the ejecta cools and the Planck function peaks at longer wavelengths.  

The opacities we infer for the \emph{individual} ejecta components are also similar to values commonly assumed in multi-component analytic light curve models. Our posterior distribution for the opacity of the blue component (visible from 1.4 to 4.4 days) ranges from $1.9^{+0.2}_{-0.2}~\mathrm{cm^2~g^{-1}}$~to $9.8^{+20.0}_{-6.6}~\mathrm{cm^2~g^{-1}}$, while our posterior distribution for the opacity of the red component (visible from 3.4 to 4.4 days) ranges from $12.1^{+24.7}_{-8.9}~\mathrm{cm^2~g^{-1}}$~to~$16.8^{+27.5}_{-14.1}~\mathrm{cm^2~g^{-1}}$ (Figure~\ref{fig:hists-opacities}). The ranges in these opacities primarily stem from the variations in the inferred $Y_e$ of this blue component at different epochs. Our blue component's opacity bears close resemblance to the purple component in the three-component models of \citet{villar17}~and \citet{cowperthwaite17}. Our red component opacity is similar to the red component opacities from those same works, as well as the red components in \citet{drout17}, \citet{kasliwal17}, and \citet{rosswog18}.

These inferred opacities are dependent on the wavelength range over which we compute Planck mean opacities; we assess this dependence here. The range over which we compute Planck mean opacities is the wavelength range over which we trust our spectral fit to yield reasonable opacities. At 1.4 and 2.4 days, motivated by the quality of the spectral fit, we exclude wavelengths $\leqslant 4500$~\AA. Imposing these cutoffs reduces the opacity by a factor of $\sim$$12$~and~$\sim$$2.5$ at 1.4 and 2.4 days, respectively. We also vary the cutoff at 1.4 and 2.4 days by $\pm1000$~\AA~to explore the sensitivity to this cutoff. At 1.4 days, setting the cutoff to a shorter (longer) wavelength leads to an increase (decrease) in the opacity by a factor of $\sim$$6$ ($\sim$$7$). At 2.4 days, the variation is less dramatic, with the opacity increasing or decreasing by a factor of $\sim$$2.5$. At 3.4 and 4.4 days, we exclude $\leqslant 6400$~\AA, reducing the opacity by a factor of $\sim$$7~\mathrm{and}~$$\sim$$3.5$~at 3.4 and 4.4 days, respectively.  Varying the cutoff at 3.4 and 4.4 days by $\pm1000$~\AA, we find a factor of $\sim$$2$ increase/decrease in the opacity at 3.4 days and a $\sim$$5-10\%$ increase/decrease at 4.4 days. In all, the 1.4 day opacity shows the largest sensitivity to the choice of wavelength cutoff, while the opacities at 2.4 and 3.4 days increase or decrease by a smaller factor of $\sim$$2 - 2.5$, and the opacity at 4.4 days varies by just $\sim$$5-10\%$. The choice of wavelength cutoff is evidently consequential, but most so at 1.4 days. We find that it is most important to exclude the poorly-modeled UV region of the spectrum in order to compute accurate opacities, with the greatest consequences at 1.4 days when the spectrum is most blue. The impact of the cutoff becomes less pronounced at later epochs. 

Finally, our resultant light curves and inferred ejecta mass depend on the functional form for the opacity provided to the light curve model. Given our inferred opacities with appropriate wavelength cutoffs, we fit the opacities with a power-law $\kappa(t) = \kappa_0 (t / ~\mathrm{1~day})^{\gamma}$~and find $\kappa_0 = 1.17 \pm 0.18~\mathrm{cm^2~g^{-1}}$ and $\gamma = 1.4 \pm 0.4$.  Our opacity evolves more steeply than that of \citet{waxman18}, who instead find $0.5 \lesssim \gamma \lesssim 1$~with $\gamma \approx 0.6$. Including later epochs in our fit could yield a similar, less steep evolution in the opacity, but would again require careful modeling of non-LTE effects (\citealt{pognan22}). Nevertheless, the fitted $t^{1.4}$~power-law opacity that we input into our light curve model accurately reproduces the light curves of the GW170817 kilonova. Interestingly, our time-dependent opacities bear some resemblance to the time-dependent Planck mean opacities of \citet{tanaka20}, which are obtained using systematic opacity calculations for elements from $Z = 26$~to~$88$. They find, for a mixture of elements with $Y_e \sim 0.1 - 0.2$~or~$Y_e \sim 0.2 - 0.3$, an opacity which trends upwards from 1 to 6 days with a sub-peak at $\sim$$2~\mathrm{days}$ (see their Figure 12). Given the uncertainties in our inferred opacities, we cannot confidently state whether there is a peak in opacity at $\sim$$2-3~\mathrm{days}$, but our results hint at such a peak. Obtaining the opacities from spectral inference at 5.4 days and beyond, with appropriate attention to non-LTE effects, could enable more robust comparison to systematic opacity calculations such as \citet{tanaka20} (see also \citealt{fontes20, kato24}). Such a comparison would also serve as a probe of atomic physics, as these calculations rely on atomic structure calculations to generate theoretical atomic line lists. 

The consistency of our opacities with the values assumed in analytic light curve models, and with the values obtained from sophisticated systematic opacity calculations, suggests that our inferred opacities are reasonable and establish a viable link between the spectral and time domains. We do not assume \textit{a priori} that tying the quantities inferred from Monte Carlo radiative transfer with \TARDIS~to a simple, analytic, gray opacity kilonova model such as that from \cite{hotokezaka20} is a valid approach. The ability to extend the opacities (and heating rates) inferred from spectral observations, sensitive only to the line-forming region and indeed only a small fraction of the ejecta given our $v_{\mathrm{inner}}$~lying above the photosphere, down to lower velocities where the bulk of the ejecta mass is contained, is not trivial. Our reasonable photometric fit demonstrates that this approach may be feasible, at least for the GW170817 kilonova.

\subsection{Physical Origin of the Ejecta}\label{ssc:disco-masses}

Modeling both spectra and light curves \textit{together} can help illuminate the physical origins of the ejecta, such as the relative contributions from dynamical or post-merger ejecta. Our spectral inference with \SPARK~provides fundamental insight into the composition of the ejecta. From our inference in the spectral-domain in \vieiratwothree, \vieiratwofour, and this work, we infer the presence of a blue ejecta component from 1.4 to 4.4 days with $Y_e \sim 0.23 - 0.31$~and specific entropy $s/k_{\mathrm{B}} \sim 13 - 18$, yielding an abundance pattern which drops below $10^{-9} - 10^{-6}$ for the lanthanides. This component could originate from the winds of a remnant hyper massive neutron star and/or a remnant accretion disk. Beginning at 3.4 days, we infer an additional red component with $Y_e \sim 0.16 - 0.21$~and~$s/k_{\mathrm{B}} \sim 18 - 22$, yielding substantial abundances of heavier elements and especially the lanthanides. These values are characteristic of the neutron-rich ejecta from a magnetized accretion disk and/or tidal ejecta. Because spectral modeling is only sensitive to the material in the line-forming region, we are unable to infer the ejecta mass of either component, nor the total ejecta mass. Indeed, we find just $\sim$$10^{-3}~M_{\odot}$~in the line-forming region at 4.4 days, when $v_{\mathrm{inner}}$~has receded considerably. From our inference in the time-domain in this work, we now find a large \textit{total} ejecta mass of $0.11^{+0.04}_{-0.02}~M_{\odot}$~between $v_{\mathrm{min}}=0.1c$ and $v_{\mathrm{max}}=0.4c$. This mass is not sensitive to the choice of $v_{\mathrm{min}}$, with the fit yielding similar masses $\pm\sim0.001~M_{\odot}$~for $v_{\mathrm{min}}$~in the range $0.05c - 0.15c$. Such a large total ejecta mass is difficult to explain without substantial contribution from post-merger ejecta sources; this provides the final insight needed to pinpoint the physical ejection mechanisms at play. 

Our light curve inference of a large total ejecta mass, on the edge of what is possible with post-merger ejecta and virtually impossible for the dynamical ejecta, implies that post-merger ejecta must contribute substantially to the total ejecta mass. We infer beginning at 3.4 days that the red component extends out to velocities of $0.35c$, above the photosphere even at the earliest times. This red component shows complete overlap in radial space with the blue component at 3.4 and 4.4 days, suggesting equipartition in masses between the two components, in the line-forming region. Indeed, a substantial red component must be present. However, this red component is not needed to explain the early spectra at 1.4 and 2.4 days. If present at those earlier times, this red component would have been outshined by a blue component that should also be substantially massive. Finally, the compositions inferred with \SPARK~are characterized by $Y_e \sim 0.23 - 0.31$, $s/k_{\mathrm{B}} \sim 13 - 18$~for the blue component and $Y_e \sim 0.16 - 0.21$, $s/k_{\mathrm{B}} \sim 18 - 22$~for the red component. These components are distinct in both electron fraction and entropy. This distinctness implies that these components must have different ejection mechanisms and could not both be accretion disk winds.

One possible physical explanation for the large total ejecta mass and distinctly two-component ejecta is the existence of a long-lived hypermassive neutron star (HMNS) remnant, following the merger. \citet{siegel14}, using 3D GRMHD simulations of a differentially rotating HMNS, find that a magnetically-driven wind from its surface can eject material at rates of $10^{-3} - 10^{-2}~M_{\odot}$~s$^{-1}$. \citet{metzger18}, with an eye to GW170817, find that a HMNS with a surface magnetic field strength of $B \approx (1-3) \times 10^{14}~\mathrm{G}$~and lifetime of $t_{\mathrm{rem}} \sim 0.1 - 1~\mathrm{s}$ before gravitational collapse to a black hole could yield a high-velocity ($v \approx 0.25 - 0.3c$) blue wind from the surface of the HMNS with a mass of $\sim$$0.02~M_{\odot}$. The rate of mass loss increases with increasing $B$~and rotation rate, due to centrifugal slinging (\citealt{metzger08}). Given the simultaneous existence of a red component, \citet{metzger18} argue for a maximum blue ejecta mass $M_{\mathrm{ej,blue}}^{\mathrm{max}} \sim 10^{-2} (t_{\mathrm{rem}} / 0.1~\mathrm{s})~M_{\odot}$. From our spectral inference, we find $\sim$$10^{-3}~M_{\odot}$~of blue material in the line-forming region at 4.4 days. This strict lower limit on the mass implies a minimum remnant lifetime of $10^{-2}$~s if the blue ejecta originates from such a HMNS remnant wind. If we instead assume that the roughly equal ratio of the red and blue components inferred from $v = 0.17c$~to $0.35c$~remains constant down to a velocity of $0.1c$, and hence that the blue component accounts for $\sim$half of the total ejecta mass we infer from the light curves, then we require a longer lifetime for the HMNS of $\sim$$0.5$~s to produce $\sim$$0.05~M_{\odot}$~of blue ejecta. More recently, using 3D GRMHD with neutrinos, \citet{kiuchi24} find that as much as $\sim$$0.1~M_{\odot}$ could be launched in this long-lived magnetar remnant scenario. \citet{curtis24}~similarly use 3D GRMHD simulations with neutrinos of a neutron star merger remnant, and find that surface winds carry $8 \times 10^{-2}~M_{\odot}$~s$^{-1}$ in a quasi-steady state, and argue that such a wind \textit{alone} could produce a blue kilonova for sufficiently long remnant lifetimes, without the need for a blue accretion disk wind. 

Furthermore, a strongly magnetized central remnant would likely be associated with a large-scale poloidal magnetic field in an accretion disk. Previous works caution that a long-lived HMNS remnant would significantly irradiate a remnant accretion disk with neutrinos, protonizing the ejecta and suppressing the synthesis of the heaviest elements (\eg, \citealt{margalit17}). However, \citet{christie19}~may provide a way to maintain the presence of redder disk winds for longer HMNS remnant lifetimes. A large-scale poloidal field associated with a strongly-magnetized HMNS remnant would encourage fast disk outflows, fast enough to escape some neutrino reprocessing, yielding a massive, fast, and neutron-rich disk ejecta (\eg, \citealt{christie19}). In contrast, a smaller-scale poloidal (or large-scale toroidal) field in the accretion disk yields less massive, slower outflows which are subject to greater neutrino reprocessing, raising the $Y_e$~(\eg, \citealt{hayashi22}). A strongly magnetized central remnant could thus also support the existence of a fast, lower-$Y_e$~outflow from the disk, explaining the red component from our spectral inference. 

Finally, we find that that we are able to accurately reproduce the observed lightcurves with $v_{\mathrm{min}} = 0.1c$~and~$v_{\mathrm{max}} = 0.4c$, though varying $v_{\mathrm{min}}$~in the range $0.05 - 0.15c$~yields fits of similar qualities and ejecta masses. Ejecta spanning $v_{\mathrm{min}}=0.05c$~to~$v_{\mathrm{max}} = 0.4c$~with a $\rho \propto v^{-4.5}$~density profile has a mass-weighted average velocity of $0.10c$, while ejecta from $v_{\mathrm{min}}=0.10c$~to~$v_{\mathrm{max}} = 0.4c$ has a mass-weighted average of $0.17c$. A blue HMNS remnant wind lies well within $0.1 - 0.4c$, but accretion disk winds may not without the support of a large-scale magnetic field supporting fast launching of these winds. For example, \citet{christie19}~find average velocities of $0.18c$~in the presence of a strong poloidal field, compared to $0.08c$~for a weak poloidal field or $0.05c$~for a toroidal field. Our ability to reproduce the light curves with $v_{\mathrm{min}} = 0.1c$~thus also supports the existence of overall fast post-merger ejecta.

In summary, the inferred composition of the ejecta from spectral modeling and the large total ejecta mass from light curve modeling suggest: (1) the existence of both red and blue components in the ejecta, (2) the existence of a long-lived hypermassive magnetar remnant, launching substantial blue winds, and (3) neutron-rich outflows from a magnetized accretion disk as the source of the red ejecta, with a potential contribution from tidal ejecta. This complete, coherent picture is painted by analyzing the kilonova in both the spectral and time domains, and reaffirms previous inferences of the presence of a HMNS magnetar remnant and a massive red wind from a magnetized accretion disk.


\section{Conclusions}\label{sec:conco}

With our tool \SPARK, we fit the optically-thick, early-time, optical/infrared spectra of the GW170817 kilonova at 1.4, 2.4, 3.4, and 4.4 days post-merger. We directly measure the element-by-element abundance pattern of the kilonova ejecta, with uncertainties. A key limitation of our spectral analysis is that our inferred inner boundary velocity lies above the photosphere and thus our inferred abundances and other ejecta parameters do not describe the entirety of the photospheric ejecta. Nonetheless, leveraging the results of parametric nuclear reaction network calculations, we turn our measurement of the kilonova ejecta's composition into a measurement of the time-dependent radioactive heating rate $\dot{q}(t)$ and the wavelength-, time-, and velocity-dependent opacity $\kappa(\lambda, t, v)$ of the ejecta. We report in detail these heating rates in Tables~\ref{tab:bestfit_single}~and~\ref{tab:bestfit_multi}, and the opacities in Tables~\ref{tab:opacities}~and~\ref{tab:opacities-blured} (Appendix~\ref{app:opacities}).

We input these opacities and heating rates into a simple kilonova light curve model, and fit the bolometric light curve of the GW170817 kilonova to infer the total ejecta mass---a quantity which cannot be obtained from spectral modeling---while accurately reproducing the observed light curves. We find a total ejecta mass of $0.11^{+0.04}_{-0.02}~M_{\odot}$~between $0.1c$~and~$0.4c$~in our light curve fit. This large ejecta mass can be explained by invoking a hypermassive magnetar remnant which survives for $\sim$$0.01 - 0.5~\mathrm{s}$ post-merger, launching as much as $\sim$$0.05~M_{\odot}$~of blue ejecta material from its surface for longer lifetimes. This component coexists with a red accretion disk wind, which may be accelerated to higher velocities by a large-scale poloidal magnetic field in the disk, associated with the magnetar. Tidal ejecta could also contribute to this red component.

Our spectrally inferred heating rates and opacities probe only the line-forming region, but extending these inferred quantities down to smaller velocities where the bulk of the ejecta is located via a simple light curve model produces an accurate fit to the photometric observations. Having established this bridge between the spectral and time domains, we put forward a coherent model for the kilonova. Our spectral inference provides the detailed composition and physical conditions of the ejecta, while our light curve modeling is required to measure the total ejecta mass. This combination thus enables us to more robustly determine the physical origins of the ejecta. Contemporaneous spectroscopic and photometric observations will be critical to fully illuminating and understanding the diverse properties of a future sample of kilonovae.

\begin{acknowledgements}

We thank the anonymous referee for their extensive and thorough comments, which have strengthened this study.

N.V.\ works in Tiohti{\'a}:ke/Mooniyang, also known as Montr{\'e}al, which lies on the unceded land of the Haudenosaunee and Anishinaabeg nations. This work made use of high-performance computing resources in Tiohti{\'a}:ke/Mooniyang and in Burnaby, British Columbia, the unceded land of the Coast Salish peoples, including the Tsleil-Waututh, Kwikwetlem, Squamish, and Musqueam nations. We acknowledge the ongoing struggles of Indigenous peoples on this land, and elsewhere on Turtle Island---their land, not settlers' land---and work towards a future marked by true reconciliation.

This work made extensive use of the \href{https://docs.alliancecan.ca/wiki/Narval/en}{\texttt{Narval}} and \href{https://docs.alliancecan.ca/wiki/Cedar}{\texttt{Cedar}} clusters of the \href{https://alliancecan.ca/en}{Digital Research Alliance of Canada} at the {\'E}cole de technologie sup{\'e}rieure and Simon Fraser University, respectively. We thank the support staff of Calcul Qu{\'e}bec in particular for their assistance at various steps in this project. 

This work made use of the \href{http://vald.astro.uu.se/~vald/php/vald.php}{Vienna Atomic Line Database (VALD)}, operated at Uppsala University, the Institute of Astronomy RAS in Moscow, and the University of Vienna. We thank Nikolai Piskunov and Eric Stempels for help in obtaining the VALD data.

This research also made use of \href{https://tardis-sn.github.io/tardis/index.html}{\TARDIS}, a community-developed software package for spectral synthesis in supernovae (\citealt{kerzendorf14}). The development of \TARDIS~received support from Github, the Google Summer of Code initiative, and from the European Space Agency's (ESA) Summer of Code in Space program. \TARDIS~is a fiscally
sponsored project of NumFOCUS. \TARDIS~makes extensive use of \href{https://docs.astropy.org/en/stable/}{\texttt{astropy}} and \href{https://pyne.io/}{\texttt{PyNE}}. We thank Andrew Fullard, Wolfgang Kerzendorf, and the entire \TARDIS~development team for their assistance and their commitment to the development and maintenance of the code. 

N.V.\ thanks Albert Sneppen, Rasmus Nielsen, and Darach Watson for their generous hospitality and generative discussions while visiting the Niels Bohr Institute in March 2024. N.V.\ also thanks the attendees of the September 2023 Radiative Transfer and Atomic Physics of Kilonovae conference and workshop in Stockholm, Sweden for productive and insightful discussions on kilonovae and their many outstanding mysteries. 

N.V.\ acknowledges funding from the Natural Sciences and Engineering Research Council of Canada (NSERC) Canada Graduate Scholarship - Doctoral (CGS-D) and the Murata Family Fellowship. J.J.R.\ acknowledges support from the Canada Research Chairs (CRC) program, the NSERC Discovery Grant program, the Canada Foundation for Innovation (CFI), and the Qu\'{e}bec Ministère de l’\'{E}conomie et de l’Innovation. 
D.H. acknowledges funding from the NSERC Arthur B. McDonald Fellowship and Discovery Grant programs and the Canada Research Chairs (CRC) program. The authors acknowledge support from the Centre de recherche en astrophysique du Québec, un regroupement stratégique du FRQNT.
M.R.D. acknowledges support from the NSERC through grant RGPIN-2019-06186, the Canada Research Chairs Program, and the Dunlap Institute at the University of Toronto. 
R.F. acknowledges support from NSERC through Discovery Grant No. RGPIN-2022-03463.
\newline
\end{acknowledgements}

\software{
\href{https://dflemin3.github.io/approxposterior/index.html}{\approxposterior}: \cite{fleming18};
\href{https://docs.astropy.org/en/stable/}{\texttt{astropy}}: \cite{astropy18}; \href{https://lscsoft.docs.ligo.org/bilby/index.html}{\texttt{bilby}}: \cite{ashton19}; \href{https://cmasher.readthedocs.io/}{\texttt{cmasher}}: \cite{velden20};
\href{https://corner.readthedocs.io/en/latest/index.html}{\texttt{corner}}: \cite{foreman-mackey16};
\href{https://dynesty.readthedocs.io/en/latest/index.html}{\texttt{dynesty}}: 
\cite{speagle20}; 
\href{https://george.readthedocs.io/en/latest/}{\texttt{george}}: \cite{ambikasaran15};
\href{https://github.com/Basdorsman/kilonova-heating-rate}{\texttt{kilonova-heating-rate}};
\href{https://redback.readthedocs.io/en/latest/}{\texttt{redback}}: \cite{sarin24a};
\href{https://tardis-sn.github.io/tardis/index.html}{\TARDIS}: \cite{kerzendorf14, kerzendorf23}
}

\bibliographystyle{apj}

\appendix{}

\section{Radioactive heating rates 
 for distinct components}\label{app:radioactive-components}

\begin{figure}[!ht]
    \centering
    \includegraphics[width=0.48\textwidth]{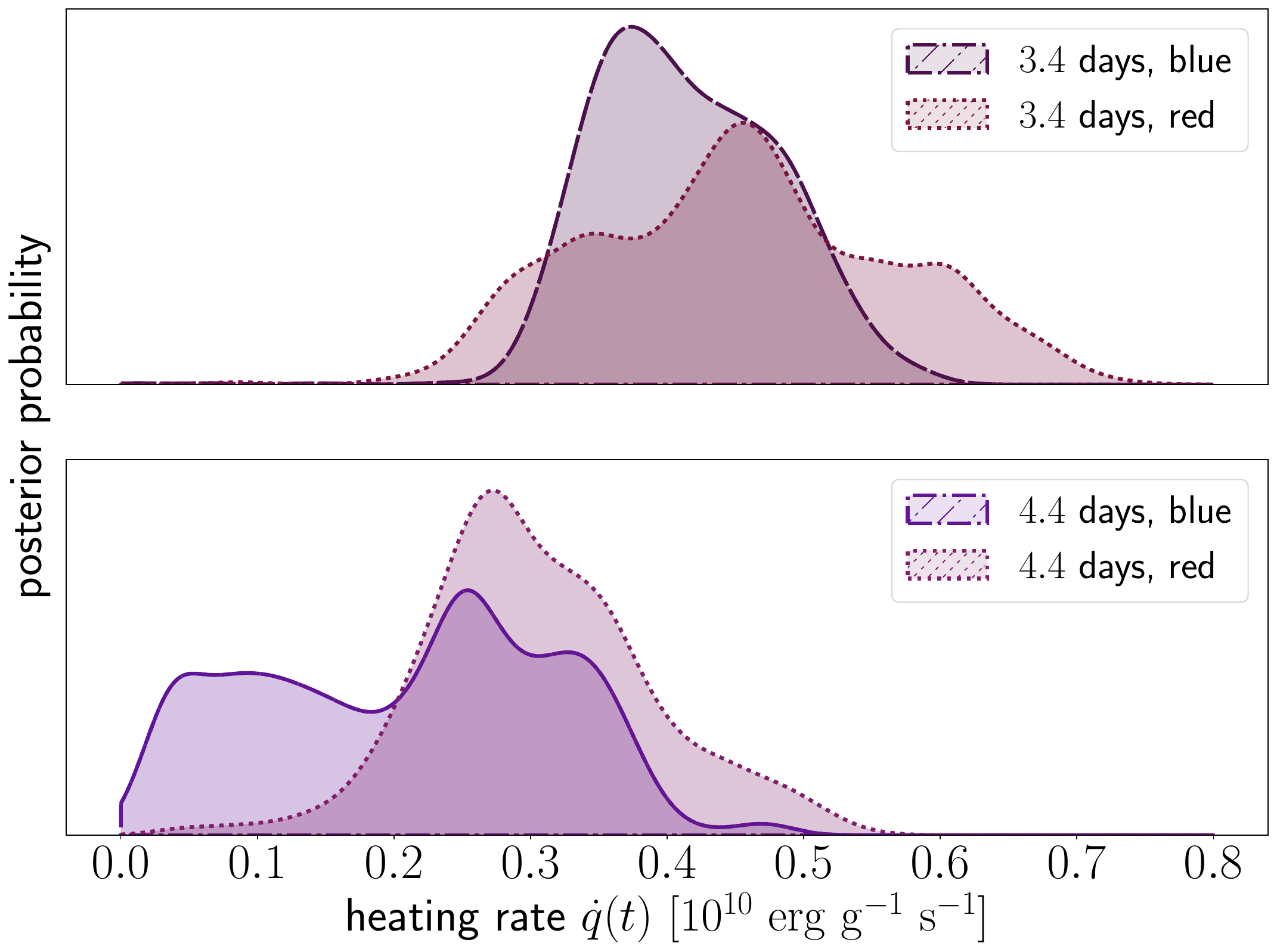}
    \caption{\textbf{Inferred posterior probability distributions of the radioactive heating rates for the distinct blue and red components at 3.4 and 4.4 days.} Heating rates are extracted from the nuclear reaction network calculations of \cite{wanajo18} given the inferred $Y_e,~v_{\mathrm{exp}},~\mathrm{and}~s$ for each component at each epoch.}
    \label{fig:hists-heatingrates-components}
\end{figure}

In Table~\ref{tab:bestfit_multi}, we report the best-fit parameters and derived radioactive heating rates $\dot{q}$ of the multicomponent fits at 3.4 and 4.4 days. For visualization, in Figure~\ref{fig:hists-heatingrates-components}, we show the posterior probability distributions for the heating rates of the distinct blue and red components at these epochs.

\section{Inferred opacities}\label{app:opacities}

In Table~\ref{tab:opacities}, we report the ingredients which go into calculations of the expansion and Planck mean opacities (velocities, densities, temperatures, and wavelength cutoffs) and the resultant opacities. At 3.4 and 4.4 days, we include opacities of both components, and the total opacity. The total opacity is obtained by mass-weighting and summing both components \textit{a priori} and computing the opacity of this mixture of two components. This approach captures the physics of reprocessing, whereby one component may absorb and reemit the emission of the other. In Table~\ref{tab:opacities-blured}, we detail the opacities of each component at 3.4 and 4.4 days. The red and blue components at 3.4 and 4.4 days are slightly offset in velocity/radius, resulting in a slightly different set of velocity bounds and densities in each component, but there is no strong evidence for a radial gradient in composition. 

\newpage

\begin{deluxetable*}{c|cccc}[ht]
\centering
\tablecaption{Ingredients in the opacity calculations and resultant opacities at each epoch. For each epoch at time $t_{\mathrm{exp}}$ post-merger/explosion: the inner and outer boundary velocities, velocity in the middle of the shell, density at this $t_{\mathrm{exp}}$ and $v_{\mathrm{middle}}$, (fixed) temperature at the inner boundary the ejecta, radiation temperature of the ejecta given the selected $T_{\mathrm{inner}}$, Planck mean opacities of the blue, red, and combined mixed ejecta as applicable, (fixed) wavelength cutoff used in calculating the preferred Planck mean opacities, and opacities with this cutoff imposed.}
\tablehead{parameter & 1.4 days & 2.4 days & 3.4 days & 4.4 days}
\startdata\tablewidth{1.0\textwidth}
$v_{\mathrm{inner}}/c$ & $0.313^{+0.013}_{-0.014}$ & $0.249^{+0.017}_{-0.032}$ & $0.213^{+0.056}_{-0.035}$ & $0.172^{+0.047}_{-0.024}$ \\
$v_{\mathrm{outer}}/c$ & $0.35$ & $0.342^{+0.047}_{-0.050}$ & $0.344^{+0.035}_{-0.038}$ & $0.337^{+0.041}_{-0.045}$ \\
$v_{\mathrm{middle}}/c$ & $0.331^{+0.006}_{-0.007}$ & $0.296^{+0.022}_{-0.024}$ & $0.278^{+0.027}_{-0.0286}$ & $0.254^{+0.025}_{-0.013}$ \\\hline
$\rho (t=t_{\mathrm{exp}}, v=v_{\mathrm{middle}})~[10^{-16}~\mathrm{g~cm^{-3}}]$ & $6.30^{+0.39}_{-0.32}$ & $1.80^{+0.52}_{-0.35}$ & $0.76^{+0.28}_{-0.18}$ & $0.46^{+0.08}_{-0.12}$ \\\hline
$T_{\mathrm{inner}}~[\mathrm{K}]$ & 5440 & 3940 & 3420 & 3330 \\
$T_{\mathrm{rad}}~[\mathrm{K}]$ & $5302^{+30}_{-34}$ & $3727^{+92}_{-94}$ & $3192^{+38}_{-55}$ & $3048^{+29}_{-79}$ \\\hline
$\kappa_{\mathrm{Planck,blu}}~[\mathrm{cm^2~g^{-1}}]$ & $24.0^{+2.9}_{-3.8}$ & $27.5^{+7.4}_{-6.2}$ & $68.7^{+112.1}_{-54.4}$ & $15.1^{+195.5}_{-7.0}$ \\
$\kappa_{\mathrm{Planck,red}}~[\mathrm{cm^2~g^{-1}}]$ & - & - & $118.9^{+144.8}_{-100.9}$ & $78.0^{+142.0}_{-66.2}$ \\
$\kappa_{\mathrm{Planck,total}}~[\mathrm{cm^2~g^{-1}}]$ & $24.0^{+2.9}_{-3.8}$ & $27.5^{+7.4}_{-6.2}$ & $65.8^{+106.9}_{-51.5}$ & $14.2^{+121.6}_{-6.1}$ \\\hline
$\lambda_{\mathrm{cutoff}}$ [\AA] & 4500 & 4500 & 6400 & 6400 \\
$\kappa_{\mathrm{Planck,blu}} (\lambda \geqslant \lambda_{\mathrm{cutoff}})~[\mathrm{cm^2~g^{-1}}]$ & $1.9^{+0.2}_{-0.2}$ & $9.7^{+5.1}_{-3.1}$ & $9.8^{+20.0}_{-6.6}$ & $4.2^{+28.9}_{-1.2}$ \\
$\kappa_{\mathrm{Planck,red}} (\lambda \geqslant \lambda_{\mathrm{cutoff}})~[\mathrm{cm^2~g^{-1}}]$ & - & - & $16.8^{+27.5}_{-14.1}$ & $12.1^{+24.7}_{-8.9}$ \\
$\kappa_{\mathrm{Planck,total}} (\lambda \geqslant \lambda_{\mathrm{cutoff}})~[\mathrm{cm^2~g^{-1}}]$ & $1.9^{+0.2}_{-0.2}$ & $9.7^{+5.1}_{-3.1}$ & $9.6^{+18.5}_{-6.4}$ & $3.9^{+17.4}_{-0.9}$
\enddata
\end{deluxetable*}\label{tab:opacities}

\begin{deluxetable*}{c|cccc}[ht]
\centering
\tablecaption{Same as Table~\ref{tab:opacities}, but detailed for the distinct blue and red components at 3.4 and 4.4 days.}
\tablehead{parameter & 3.4 days, blue & 3.4 days, red & 4.4 days, blue & 4.4 days, red}
\startdata\tablewidth{1.0\textwidth}
$v_{\mathrm{inner}}/c$ & $0.213^{+0.056}_{-0.035}$ & $0.232^{+0.038}_{-0.027}$ & $0.172^{+0.047}_{-0.024}$ & $0.184^{+0.046}_{-0.022}$ \\
$v_{\mathrm{outer}}/c$ & $0.344^{+0.035}_{-0.038}$ & $0.334^{+0.037}_{-0.022}$ & $0.329^{+0.056}_{-0.060}$ & $0.337^{+0.041}_{-0.045}$ \\
$v_{\mathrm{middle}}/c$ & $0.279^{+0.030}_{-0.030}$ & $0.283^{+0.024}_{-0.022}$ & $0.250^{+0.025}_{-0.016}$ & $0.260^{+0.020}_{-0.018}$ \\\hline
$\rho (t=t_{\mathrm{exp}}, v=v_{\mathrm{middle}})~[10^{-16}~\mathrm{g~cm^{-3}}]$ & $0.76^{+0.31}_{-0.202}$ & $0.72^{+0.20}_{-0.16}$ & $0.49^{+0.11}_{-0.12}$ & $0.43^{+0.11}_{-0.09}$ \\\hline
$T_{\mathrm{inner}}~[\mathrm{K}]$ & 3420 & 3420 & 3330 & 3330 \\
$T_{\mathrm{rad}}~[\mathrm{K}]$ & $3192^{+68}_{-55}$ & $3235^{+39}_{-40}$ & $3060^{+104}_{-91}$ & $3064^{+85}_{-68}$ \\\hline
$\kappa_{\mathrm{Planck}}~[\mathrm{cm^2~g^{-1}}]$ & $68.7^{+112.1}_{-54.4}$ & $118.9^{+144.8}_{-100.9}$ & $15.1^{+195.5}_{-7.0}$ & $78.0^{+142.0}_{-66.2}$ \\\hline
$\lambda_{\mathrm{cutoff}}$ [\AA] & 6400 & 6400 & 6400 & 6400 \\
$\kappa_{\mathrm{Planck}} (\lambda \geqslant \lambda_{\mathrm{cutoff}})~[\mathrm{cm^2~g^{-1}}]$ & $9.8^{+20.0}_{-6.6}$ & $16.8^{+27.5}_{-14.1}$ & $4.2^{+28.9}_{-1.2}$ & $12.1^{+24.7}_{-8.9}$
\enddata
\end{deluxetable*}\label{tab:opacities-blured}


\end{document}